\documentstyle[12pt]{article}
\newcommand{\k}{\mbox{\boldmath$k$}}

\newcommand{\q}{\mbox{\boldmath$q$}}
\newcommand{\Q}{\mbox{\boldmath$Q$}}

\newcommand\up{\uparrow}
\newcommand\down{\downarrow}
\newcommand\dwon{\downarrow}
\newcommand{\lspin}{\mbox{\boldmath$S$}}
\newcommand{\coskx}{\cos  k_{x}a}
\newcommand{\cosky}{\cos  k_{y}a}

\newcommand{\spin}{\mbox{\boldmath$\sigma$}}

\newcommand{\etal}{\it {et al.}}
\newcommand{\freq}{\omega_{l}}
\newcommand{\jpsj}{J.Phys.Soc.Jpn.}

\newcommand{\prl}{ Phys. Rev. Lett.}
\newcommand{\prb}{ Phys. Rev.{\bf  B}.}
\newcommand{\pr}{ Phys. Rev.}
\newcommand{\prog}{Prog.Theoret.Phys}
\title{ Spin Fluctuations in \\
Magnetically Coupled
Bi-layer Cuprates} 
\author{JunIchiro KISHINE\thanks{kishine@watson.phys.s.u-tokyo.ac.jp}\\ 
\it Department of Physics, Faculty of Science, \\
\it University of Tokyo, 7-3-1 Hongo, Tokyo113.}
\date{(Received April 10, 1996)\\
Submitted to Journal of the Physical Society of Japan}
\begin{document}
\maketitle
\begin{abstract}
We propose a possible mechanism of pseudo spin gap anomaly(PSGA)
in magnetically coupled bi-layer cuprates without any fermion pairing instability.
In our proposal PSGA does not necessarily require the spin-charge
separation  or the breakdown of the Fermi liquid description of 
a normal state of the cuprate superconductor.

The low energy magnetic excitations are mainly governed by the
{\it itinerant nature of the intra-layer system} and 
{\it the inter-layer antiferromagnetic coupling}.
No matter how 
weak the bare inter-layer coupling is, it can be dramatically enhanced due to 
the intra-layer spin fluctuations.
As the temperature decreases  near the antiferromagnetic phase boundary 
the strongly
enhanced inter-layer correlation induces the inter-layer 
particle-hole exchange scattering processes
that tend to enhance  the inter-layer spin singlet formation and
kill the triplet formation.
We propose that the coupling of spin fluctuations on the adjacend layers
via the strong repulsive interaction
 between parallel spins travelling on each layer give rise to the dynamical screening
effects.
As a result the low energy part of the spin excitation spectrum is
strongly suppressed as the temperature decreases near the antiferromagnetic phase boundary.
We ascribe PSGA to this dynamical screening effects. \\\\\\
{\bf KEYWORDS}: magnetically coupled bi-layer, itineracy, spin fluctuations, spin gap 
\end{abstract}
\pagebreak
\section{Introduction} 
\baselineskip 15pt
Since the discovery of the  high $T_c$ cuprate superconductor in 1986,  puzzling features 
in their normal state due to strong correlation\cite{Anderson}, especially the spin dynamics of 
them\cite{kampf} 
have provoked a great deal of controversy. 
In the high $T_c$ superconductor, a superconducting phase lies just near the Mott
insulating
phase where only the spin degree of freedom survives. 
Thus even in a metallic phase 
the antiferromagnetic spin fluctuations play essential roles to understand 
the physical nature of these materials.
Especially the spin dynamics of bi-layer cuprates have  recently 
attracted a lot of 
attention,  since the so-called pseudo spin gap anomaly(PSGA), 
that is one of the most serious problem in this field, 
seems to be peculiar to  bi-layer cuprates. 

PSGA was first observed  by Yasuoka\cite{typeA6.6-0} as
the phenomena that  NMR longitudinal
relaxation rates, $1/T_{1}T$ deviate from Curie-Weiss like temperature
dependence well above the superconducting transition temperature and show
a broad peak around the so-called spin gap temperature  before 
it decreases
smoothly down through  $T_{c}$ with slight change of slope. 
Up to now the clear signature of PSGA in $1/T_{1}T$ can be observed in 
YBa$_{2}$Cu$_{3}$O$_{6+x}$
\cite{typeA6.6-0,typeA6.6-1,typeA6.6-2,typeA6.6-3,typeA123 6.6-4,typeA Y6.6 and 
La123,typeA1236.9-1,typeA123 6.98-1,typeA123 dope dep on spin gap} and ,
YBa$_{2}$Cu$_{4}$O$_{8}$\cite{typeA248-1,typeA248-2,typeA248-3}, 
Y$_{2}$Ba$_{4}$Cu$_{7}$O$_{15}$\cite{typeA24715-1},  
LaBa$_{2}$Cu$_{3}$O$_{y}$\cite{typeALa123} , and  
Bi$_{2}$Sr$_{2}$CaCu$_{2}$O$_{y}$\cite{typeA Bi2212,typeA Bi2223}
that 
are all 'bi-layer' cuprates where there are 
two equivalent CuO$_{2}$ layer in  a unit cell.
Furthermore in the lightly doped YBa$_{2}$Cu$_{3}$O$_{6+x}$, 
the inelastic  neutron scattering cross sections
deviate from a linear $\omega$ dependence in the low energy region and
are strongly suppressed above the superconducting transition 
temperature\cite{RossatMignod1,RossatMignod2,RossatMignod3,Shirane6.6Gap}.

On the other hand in typical 'mono-layer' cuprate 
La$_{2-x}$Sr$_{x}$CuO$_{4}$,
$1/T_{1}T$ begin to  deviate slightly from Curie-Weiss law
below  $T\sim200$K\cite{Yasuoka}, but  do not show clear peak structure 
above $T_{c}$. 
As for the neutron experiments on LSCO system, only recently
an overall energy spectrum of magnetic fluctuations was reported  \cite{directLa} in 
homogeneous single crystal of La$_{1.85}$Sr$_{0.15}$CuO$_{4}$.
They first reported the gap formation in LSCO system {\it below} $T_{c}$
and succeeded in detecting  the superconducting phase of this system.
However they didn't detect the pseudo-spin gap anomaly in this compound.
Furthermore the single layer over-doped cuprate Tl$_{2}$Ba$_{2}$CuO$_{y}$ 
 exhibits a very broad peak in both of $1/T_{1}T$ and $1/T_{2}$ slightly above $T_{c}$. 
As  Yasuoka pointed out\cite{typeALa123,typeB-1,typeB-2}, 
this behavior should be atributed to the band effects and
cannnot be regarded as the spin-gap behavior.
These experimental facts indicate
that PSGA is  peculiar to  bi-layer cuprates.

The most important feature of PSGA is that 
 the strong suppression of the low energy spin excitations
 does not accompany
the clear gap like structure in the low energy charge excitations.
By reason of this fact,  the spin gap in high $T_{c}$ superconductor
is called the {\it pseudo}  gap.
The suppression of the spin excitations clearly indicates that
{\it the singlet pair formation is favored and as a result
the magnetic degree of freedom tends to be killed}\cite{bilayerHeisenberg}.
Therefore to understand PSGA phenomena we have to seek some mechanism  
that favors the singlet pair formation without 
any suppression of the charge excitations.

From theoretical viewpoints there are two fixed points to describe 
low energy excitations in a normal state of 
the cuprate superconductor.
One  fixed point is the  Fermi liquid fixed point where
the low energy excitations are described in terms of  quasi-paricle excitations.
If we try to explain PSGA within this framework, we have to 
seek a mechanism that enhances the singlet pair 
formation
{\it without any pairing instability} since  pairing instability in a quasi-particle system
 means the superconducting transition
where both of the spin and charge degree of freedom condensate into pair.
Another fixed point is the non-Fermi liquid picture
with spin charge separation
where the low energy excitations are spin $1/2$ chargeless fermions(spinons)
and charged spinless bosons(holons)\cite{Anderson}.
In this picture  PSGA is ascribed to the condensation of the
spinons into the singlet pair. 

The Hubbard model is
the simplest model to describe the effects of the strong correlation
within the Fermi liquid picture.
In the Hubbard model, we take into account the electronic correlations
by including the on-site repulsion, $U$, between two electrons with opposite spins.
Then the simplest scheme to describe the collective spin excitations is
 the random phase approximation(RPA), namely paramagnon 
theory\cite{Paramagnon}.
Apart from PSGA, quantitative behavior of  dynamical spin fluctuations can well 
be described 
within the simple RPA\cite{BulutScalapinoNMR,BulutScalapinoNeutron,TewordtRPA}.
We can go beyond RPA by treating the feedback effects of 
the spin fluctuations on the irreducible polarization in a self
consistent manner\cite{SCRoriginal,kawabata,SCR}.
More sophisticated treatment based on Baym-Kadanoff formalism
that treats the exchange of fluctuations in the particle-hole and
particle-particle channels in a self consistent manner has been 
proposed\cite{BickersScalapino,SereneHess}. 
Phenomenological approach to the antiferromagnetic spin fluctuations
proposed by Millis, Monien and Pines\cite{MMP}  also lays its microscopic
foundation on the Fermi liquid picture. 
 However it seems difficult to explain PSGA  based  only on the Hubbard model,
since {\it the Hubbard interaction never distinguishes the singlet and triplet pair
selectively}. In the Hubbard model there appear only one scattering channel between
quasi-particles with opposite spins  and as a result it cannot
describe the tendency of the enhanced singlet.

On the other hand the non-Fermi liquid picture with spin-charge separation
stems from the $t$-$J$ model that can be regarded as the strong coupling limit 
of the Hubbard model\cite{ZhangRice}.
This model can give magnetic phase diagram within the mean field level\cite{Fukuyama}.
Recently the extention  of the $t$-$J$ model to 
the bi-layer cuprates was also reported\cite{LercherWheatley,NKF,UbbensLee2}.
In the $t$-$J$ model the presence of the spin 
dependent interaction is essential. It produces three fundamental scattering channels,
one  with spin-flip and two  without spin-flip.
These scattering channels can be decomposed into the
singlet and triplet channel and spin dependent nature of 
scattering process can well be described. 
Concerning PSGA there have been some proposals mainly from the gauge field theoretical
approach based on the $t$-$J$ model\cite{IoffeLarkin,Nagaosa,UbbensLee1,AltshulerIoffe,UbbensLee2,WenLee}.
Altshuler and Ioffe\cite{AltshulerIoffe} first pointed out the importance of
bi-layer correlation.
They 
claimed that if the spins on each layer form a spin liquid state, this 
transition can be
described as a spinon pairing on adjacent layers, that leads to the spin-gap anomaly.
As for the spinon condensation into the singlet pair,
Ubbens and Lee claimed\cite{UbbensLee1} that $U(1)$ gauge fluctuations tend to
destroy the intra-layer spinon pairing, although 
it is less effective in destroying the inter-layer pairing\cite{UbbensLee2}.
However recently Wen and Lee\cite{WenLee} reported that the spin gap phase is best 
understood as the intra-layer staggered flux phase where the $SU(2)$ symmetry is 
preserved away from half filling.
Therefore within the framework of 2-dimensional gauge theory
there still exist some ambiguity on whether PSGA is intrinsic 
to bi-layer structure
or not. 

However it remains an unsettled question 
whether the low energy excitations in cuprate can be described
as the Fermi liquid or not.  
In the present paper, we should like to explore a different possibility to understand PSGA
{\it without any fermion condensation},
which has never been examined.
As stated above to understand PSGA the existence of spin dependent interaction
is essentially important.  However this does not necessary mean the spin-charge separation
really takes place.
Motivated by this consideration, we try to understand PSGA without breakdown of the  
Fermi liquid description.
To compensate the drawback of the Hubbard model that
it cannot distinguish the singlet pair formation selectively, 
we consider the {\it magnetically coupled bi-layer} cuprates and
try to understand the PSGA without any 
pairing instability.
Our motivation was first provoked by the idea by Ioffe {\it et.al}\cite{Ioffe},
where the strongly developed inter-layer 
spin fluctuations dramatically enhance the inter-layer interaction.
In the present paper we treat the intra-layer
spin fluctuations based on the Hubbard model.
As the temperature decreases, the developed intra-layer spin fluctuations  
dramatically enhance the
inter-layer magnetic correlation near the magnetic phase boundary.
Therefore no matter how weak the bare inter-layer magnetic coupling is, it can
affect the whole structure of the magnetic excitations of the system
at the low temperature near the magnetic phase boundary.

Just near the antiferromagnetic phase boundary,
the enhanced inter-layer correlations strongly induce
the inter-layer particle-hole exchange scattering processes.
In this paper we show that due to these scattering processes
the inter-layer singlet formation is strongly enhanced
while the triplet formation is strongly suppressed.
In other words interaction between parallel spins becomes strongly repulsive. 
As a result we can ascribe the
PSGA to the dynamical screening  of the low energy spin excitations 
due to this repulsion.
Actually we can show that as the temperature decreases
and the intra-layer spin fluctuations develop,
the imaginary parts of the whole susceptibility
reveal apparent gap-like structure in the low energy region.
We  can assign the PSGA to this mechanism.

The outline of this paper is as follows.
In section 2, we present the model 
and formalism.
In section 3, we present how dramatically the inter-layer bare interaction
is enhanced by the intra-layer spin fluctuations. We pay our attention 
to the spin rotational invariance relation.
In section 4, we investigate the inter-layer particle-hole exchange scattering processes
driven by the enhanced inter-layer intetaction in some detail.
In section 5,
we show that the coupling of spin fluctuations on the adjacend layers
via the inter-layer processes give rise to the dynamical screening
effects.
In section 6,
we present the
relations between microscopic spin correlations obtained in the present approach and
experimentally observed quantities in magnetically coupled bi-layer cuprates.
In section 7,
We present the numerical results on the analytically continued dynamical 
susceptibilities
and show that our proposal can actually be realized.

\section{Model and Formalism}
In this section we present our model and basic formalism to treat the spin
fluctuations in magnetically coupled bi-layer.
We consider YBa$_{2}$Cu$_{3}$O$_{6+x}$ (Y123)
compounds as  representatives of
the bi-layer family.
As shown in Fig.~1.  there are a pair of equivalent  CuO$_{2}$ layers per unit cell
that  are separated by the charge reservoir layers. We ignore a small orthorhombic distortion
of the CuO$_{2}$ lattice.
The distance between the adjacent
CuO$_{2}$ layers is approximately $c^{*}\sim 3.2 \rm \AA$. On the other hand 
the distance
between the pair of layers is approximately $c\sim 8.2 \rm \AA$. 

\subsection{Model Hamiltonian}
We start with the Hubbard Hamiltonian with small direct inter-layer hopping
\begin{equation}
{\cal{H}}=
\sum_{i,j}\sum_{\sigma}\sum_{n}
t_{ij}c_{n,j,\sigma}^{\dagger}c_{n,i,\sigma}
+t_{\perp}\sum_{\sigma}\sum_{n\neq m}
c_{n,i,\sigma}^{\dagger}c_{m,i,\sigma}
+U\sum_{i}\sum_{n}n_{n,i,\up}n_{n,i,\down},
\end{equation}
where $c_{n,i,\sigma}^{\dagger}$ ($c_{n,i,\sigma}$)
is a creation (annihilation) operator of an itinerant carrier belonging to the
$i$-th Cu site of the $n$-th layer ($n=1,2$) with the spin projection $\sigma$
and $n_{n,i,\sigma}=c{_{n,i,\sigma}}^{\dagger}c_{n,i,\sigma}$.
Furthermore $t_{ij}$ denotes the intra-layer hopping integral between the $i$-$j$ sites
and  $t_{\perp}$ denotes the inter-layer hopping integral.
As for the intra-layer on-site Hubbard repulsion, $U$,
 we assume the intermediate coupling scheme,
namely $U \sim 3t$, where $t$ is the hopping integral 
between the nearest neighbor sites.
On the other hand,  we can safely assume, $t_{\perp}\ll U$ and therefore 
we can treat the direct inter-layer hopping process in a $t$-$J$ like manner.
Thus we obtain
\begin{equation}
{\cal{H}}={\cal{H}}_{//}+{\cal{H}}_{\perp},\label{eqn:effective}
\end{equation}
where 
\begin{equation}
{\cal{H}}_{//}=\sum_{i,j}\sum_{\sigma}\sum_{n}
t_{ij}c_{n,j,\sigma}^{\dagger}c_{n,i,\sigma}
+U\sum_{i}\sum_{n}n_{n,i,\up}n_{n,i,\down},
\end{equation}
denotes the intra-layer Hubbard Hamiltonian
and 
\begin{equation}
{\cal{H}}_{\perp}=t_{\perp}\sum_{\sigma}\sum_{n\neq m}
(1-n_{n,i,-\sigma})c_{n,i,\sigma}^{\dagger}c_{m,i,\sigma}(1-n_{m,i,-\sigma})
+2J_{0\perp}\displaystyle{\sum_{i}}\displaystyle{\sum_{m \neq n}}
\lspin_{m,i}\cdot\lspin_{n,i},
\end{equation}
is the  $t$-$J$ like counterpart that gives rise to
the inter-layer magnetic coupling.
The spin fluctuation operator is introduced by
\begin{eqnarray}
\lspin^{(n)}_{i}
={1\over 2}\sum_{\alpha\beta}
c^{\dagger}_{n,i,\alpha}{\spin}_{\alpha\beta}
c_{n,i,\beta},
\end{eqnarray}
where $\spin$ denotes usual Pauli matrices and we set $\hbar=1$.
We note that 
the hopping term in ${\cal{H}}_{\perp}$ depends on the carrier density
and near the antiferromagnetic phase boundary it is reduced, while the
magnetic coupling is unlikely to be sensitive to the doping. 
Furthermore, as will be shown
later, near the phase boundary the inter-layer coupling is dramatically enhanced
by the intra-layer spin fluctuations.
Based on this fact, in ${\cal{H}}_{\perp}$
we keep only the magnetic term.

Thus we are lead to the assumption that  
the spin fluctuations in  bi-layer cuprates can well
be described by the effective Hamiltonian
\begin{equation}
{\cal{H}}_{//}=\sum_{i,j}\sum_{\sigma}\sum_{n}
t_{ij}c_{n,j,\sigma}^{\dagger}c_{n,i,\sigma}+
U\sum_{i}\sum_{n}n_{n,i,\up}n_{n,i,\down},
\end{equation}
and
\begin{equation}
{\cal{H}}^{eff}_{\perp}
=2J_{0\perp}\displaystyle{\sum_{i}}\displaystyle{\sum_{m \neq n}} 
\lspin_{m,i}\cdot\lspin_{n,i}.
\end{equation}
By taking the lattice Fourier transform 
\begin{equation}
c_{n,i,\sigma}={1\over \sqrt{N}}\sum_{\footnotesize\k}{\rm e}^{i\k\cdot{\bf i}}c_{n,{\footnotesize\k},\sigma}
,\end{equation}
the effective Hamiltonian can be written by
\begin{eqnarray}{\cal{H}}_{//}=\displaystyle{\sum_{\footnotesize\k}\sum_{\sigma}\sum_{n}}
\varepsilon_{n,{\footnotesize\k}}c_{n,{\footnotesize\k},\sigma}^{\dagger}c_{n,{\footnotesize\k},\sigma}\nonumber\\
+{U\over N}\sum_{{\footnotesize\k},{\footnotesize\k}',\q}\sum_{n=1,2}
c_{n,{\footnotesize\k+\q},\up}^{\dagger}c_{n,{\footnotesize\k},\up}
c_{n,{\footnotesize\k'-\q},\down}^{\dagger}c_{n,{\footnotesize\k}',\down},
\end{eqnarray}
and
\begin{eqnarray}
{\cal{H}}^{eff}_{\perp}
=2J_{0\perp}\displaystyle{\sum_{\footnotesize\q}}\displaystyle{\sum_{m \neq n}} 
\lspin_{m}(\q)\cdot\lspin_{n}(-\q).
\end{eqnarray}
where $N$ is the number of the lattice sites and
\begin{eqnarray}
\lspin_{n}(\q)
={1\over 2}\sum_{\footnotesize\k}\sum_{\alpha\beta}
c^{\dagger}_{n,{\footnotesize\k+\q},\alpha}{\spin}_{\alpha\beta}
c_{n,{\footnotesize\k},\beta},
\end{eqnarray}
denotes the spin fluctuation in the momentum space.
Here we note the momentum $\k$ and $\q$ are the {\it two-dimensional} momentum 
since in our model the
itinerant carriers are confined into a single plane.
In the kinetic term, 
we have included the hopping process to the second nearest neighbor sites 
\begin{eqnarray}
\varepsilon_{1,\footnotesize\k}=\varepsilon_{2,\footnotesize\k}
\equiv\varepsilon_{\k}
=-2t(\coskx+\cosky)-2t'\coskx\cosky\nonumber\\
-2t''(\cos 2k_{x}a+\cos 2k_{y}a) ,
\label{eqn:band}
\end{eqnarray}
where we put $t'=-t/5$ and $t''=t/4$
to reproduce the Fermi contour of  YBCO system\cite{NKF,HopInte}.

Now  we briefly comment on the experimental evidences for the inter-layer magnetic coupling. 
It is  well established  that  there remains the strong intra-layer antiferromagnetic
spin-spin coupling which is estimated as $J_{0//}=80^{+60}_{-30}$ meV from the neutron
experiments\cite{0.3SpinWaveAnalysis}.
On  the other hand, although existence of the inter-layer antiferromagnetic 
spin-spin interaction has been widely accepted, question about the magnitude of this
interaction had been under debates.
 Concerning this issue, the early neutron scattering experiment 
gave very weak strength of inter-layer 
magnetic coupling $J_{0\perp}\sim 0.06 J_{0//}$. 
This estimation was based on the fact that 
 the expected
optical gap of spin wave excitation, $2\sqrt{J_{0//}J_{0\perp}}$,  
in an insulating  phase
had never been observed
even up to 60 meV\cite{ShiraneLowerLimitOfLambda}.
However recently it was reproted that mid-infrared transmission and reflection mesurements
picked up the optical branch at 178.0 meV\cite{midinfra} and therefore $J_{0\perp}$
can be estimated as $J_{0\perp}\sim 0.55 J_{0//}$. 
Moreover recent NMR cross-relaxation mesurements of  
Y$_{2}$Ba$_{4}$Cu$_{7}$O$_{15}$
\cite{typeA24715-1}  also suggest that the inter-layer spin-spin 
coupling  can reach  $J_{0\perp}\sim 0.25 J_{0//}$.
Tus we can say that the importance of the inter-layer magnetic   coupling
should be duely recognized to understand the magnetic properties of
bi-layer cuprates.

\subsection{Dynamical Spin Susceptibility}

In the case of magnetically coupled bi-layer, the dynamical spin susceptibility can be written   
in the thermal Green's function formalism as
\begin{eqnarray}
\chi{^{\alpha\beta}_{mn}}(\q;i\freq)
&=&
\int_{0}^{1/T}d\tau e^{i\freq\tau}
<T_{\tau}[S^{\alpha}_{m}(\q,\tau)S^{\beta}_{n}(-\q,0)]> \nonumber\\
&=&{1\over 4} \sum_{\mu\nu\lambda\rho} 
\sigma^{\alpha}_{\mu\nu}
\Gamma^{\mu\nu;\lambda\rho}_{mn}(\q;i\freq)
\sigma^{\beta}_{\lambda\rho},
\end{eqnarray}
where 
\begin{equation}
\Gamma^{\mu\nu;\lambda\rho}_{mn}(\q;i\freq)=
\int_{0}^{1/T}d\tau e^{i\freq\tau}
{1\over N}\sum_{\footnotesize\k,\k'}
<T_{\tau}[c^{\dagger}_{m,{\footnotesize\k}',\mu}(\tau)c_{m,{\footnotesize\k'+\q},\nu}(\tau)
c^{\dagger}_{n,{\footnotesize\k+\q},\rho}(0)c_{n,{\footnotesize\k},\lambda}(0)]>,
\end{equation}
denotes the spin dependent polarization function. 
Now $m,n$ are layer indices,  $\alpha,\beta=+,-,z$ are the spin indices, 
$\mu,\nu,\lambda,\rho=\uparrow,\downarrow$ are the spin projection, 
and $\freq=2\pi  T l$ is a bosonic Matsubara frequency.
Furthermore 
\begin{eqnarray}c^{\dagger}_{n,{\footnotesize\k},\alpha}(\tau)
= \exp(\tau{\cal H}) c^{\dagger}_{n,{\footnotesize\k},\alpha}(0) \exp(-\tau{\cal H})
\end{eqnarray}
represent imaginary time dependent creation operator
 where we set $\hbar=1$ and $k_{B}=1$.
The thermal average
$<\cdots>\equiv {\rm Tr}( {\rm e}^{-{\cal H}/ T}\cdots)/{\rm Tr}{\rm e}^{-{\cal H}/ T}$
is taken under the full Hamiltonian.
Due to the  symmetry under exchange of layers,   
the spin susceptibility $\chi^{\alpha\beta}_{mn}$
has only two 
independent components with respect to $m,n$:
the intra-layer ({\it diagonal}) correlation $\chi^{\alpha\beta}_{11}
=\chi^{\alpha\beta}_{22}
\equiv \chi^{\alpha\beta}_{//}$
and the inter-layer ({\it off-diagonal}) correlation
$\chi^{\alpha\beta}_{12}
=\chi{^{\alpha\beta}_{21}}^{*}
\equiv \chi^{\alpha\beta}_{\perp}$.
Furthermore each element has two independent components in spin space:
transverse counterpart $\chi^{+-}_{mn}$ and 
longitudinal counterpart $\chi^{zz}_{mn}$, which are given by 
\begin{eqnarray}
\left\{ \begin{array}{c}
\chi^{+-}_{mn}(\q;i\freq)=\Gamma^{\up\down;\down\up}_{mn}(\q;i\freq),\\\\
\chi^{zz}_{mn}(\q;i\freq)={1\over 4}
[\Gamma^{\up\up;\up\up}_{mn}(\q;i\freq)+\Gamma^{\down\down;\down\down}_{mn}(\q;i\freq)
-\Gamma^{\up\up;\down\down}_{mn}(\q;i\freq)-\Gamma^{\down\down;\up\up}_{mn}(\q;i\freq)].
\end{array}\right.\label{eqn:pol for zz}
\end{eqnarray}
Throughout this paper we consider the paramagnetic pahse and 
therefore the dynamical spin susceptibilities
{\it must} satisfy the rotational symmetry relation in spin space;
\begin{equation}
2\chi^{zz}_{mn}(\q;i\freq)=\chi^{+-}_{mn}(\q         ;i\freq).\label{eqn:RotationalSymmetry}
\end{equation}
We should always check this relation when we carry the theoretical investigation
a stage further.
 Since we assumed there is no inter-layer carrier hopping, 
the non-interacting counterpart
can simply be written as
\begin{eqnarray}
\chi_{0}{^{+-}_{mn}}(\q         ;i\freq)
=2\chi_{0}{^{zz}_{mn}}(\q         ;i\freq)
=\delta_{mn}\chi_{0}(\q;i\freq),
\label{eqn:bare poralization}         
\end{eqnarray}
where
\begin{eqnarray}
\chi_{0}(\q;i\freq)=-{T\over N}\sum_{i\varepsilon_{n}}\sum_{\footnotesize\k}
{\cal G}_{0}(\k+\q;i\varepsilon_{n}+\freq)
{\cal G}_{0}(\k;i\varepsilon_{n})\\\nonumber=
{1\over 2N}\sum_{\k}
{
\Lambda_{\footnotesize\k,\q}(T)
(\xi_{\footnotesize\k+\q}-\xi_{\footnotesize\k})
\over 
\freq^{2}+(\xi_{\footnotesize\k+\q}-\xi_{\footnotesize\k})^{2}
\ 
}.\label{eqn:barepol}
\end{eqnarray}
Now
\begin{equation}
{\cal G}_{0}(\k;i\varepsilon_{n})={1\over i\varepsilon_{n}-\xi_{\footnotesize\k}},
\end{equation}
is a Green's function for a free carrier where
$\varepsilon_{n}=(2n+1)\pi T$ is a fermionic Matsubara frequency.
The thermal extinction factor  is given by
\begin{equation}
\Lambda_{\footnotesize\k,\q}(T)=\tanh(\displaystyle{\xi{_{\footnotesize\k+\q}}\over 2T})-
\tanh(\displaystyle{\xi_{\footnotesize\k} \over 2T}),\label{eqn:extinc}
\end{equation}
where
$
\xi_{\footnotesize\k}=\varepsilon_{\footnotesize\k}-\mu. 
$

First we consider the intra-layer spin fluctuations driven  by the
Hubbard interaction within the frame work of  random phase approximation 
(RPA).
Taking into the effects of the interaction,
instead of (\ref{eqn:barepol}),  the dynamical susceptibility is rewritten in a form,
\begin{equation}
\chi^{\alpha\beta}(\q;i\freq)=-{T\over N}\sum_{k}
{\cal G}_{0}(\k+\q;i\varepsilon_{n}+\freq)
\gamma^{\alpha\beta}(\q;i\freq)
{\cal G}_{0}(\k;i\varepsilon_{n}),
\end{equation}
where $\gamma^{\alpha\beta}(q)$ denotes the RPA triangle vertex.
Usual RPA for Hubbard model\cite{Paramagnon}, 
 gives 
\begin{eqnarray}
\left\{\begin{array}{c}
\gamma^{+-}(\q;i\freq)=
\displaystyle{1\over 1-U\chi_{0}(\q;i\freq)},\\\\
\gamma^{\sigma\sigma}(\q;i\freq)=
\displaystyle{1 \over 1-U^{2}\chi_{0}(\q;i\freq)^{2}},\\\\
\gamma^{\sigma,-\sigma}(\q;i\freq)=
-\displaystyle{U\chi_{0}(\q;i\freq)\over 1-U^{2}\chi_{0}(\q;i\freq)^{2}}.
\end{array}\right.
\end{eqnarray}
These vertices satisfy the spin-rotational invariance relation 
\begin{eqnarray}
\gamma^{\sigma\sigma}(\q;i\freq)-
\gamma^{\sigma,-\sigma}(\q;i\freq)=
\gamma^{+-}(\q;i\freq)\equiv\gamma(\q;i\freq),
\end{eqnarray}
where $\gamma(\q;i\freq)\equiv [1-U\chi_{0}(\q;i\freq)]^{-1}$ is a Stoner factor.
These vertices give the dynamical spin susceptibilities;
\begin{eqnarray}
\left\{\begin{array}{c}
\chi^{+-}(\q;i\freq)=
\chi_{0}(\q;i\freq) \gamma^{+-}(\q;i\freq),\\\\
\chi^{\sigma\sigma}(\q;i\freq)=
\chi_{0}(\q;i\freq) \gamma^{\sigma\sigma}(\q;i\freq),\\\\
\chi^{\sigma,-\sigma}(\q;i\freq)=
\chi_{0}(\q;i\freq) \gamma^{\sigma,-\sigma}(\q;i\freq),\\\\
\chi^{zz}(\q;i\freq)=\displaystyle{1\over 4}\sum_{\sigma}[\chi^{\sigma,\sigma}(\q;i\freq)-\chi^{\sigma,-\sigma}(\q;i\freq)].
\end{array}\right.
\end{eqnarray}
where we used the relation (\ref{eqn:pol for zz}).
We note that in this step the spin rotational symmetry  relation 
\begin{equation}
\chi^{+-}(\q;i\freq)=2\chi^{zz}(\q;i\freq)\equiv\chi(\q;i\freq),
\end{equation} 
is satisfied where
we define the RPA spin susceptibility
\begin{equation}
\chi(\q;i\freq)={\chi_{0}(\q;i\freq) \over 1-U\chi_{0}(\q;i\freq)}.\label{eqn:intraRPA}
\end{equation} 

\section{Enhanced Inter-layer Interaction}
The fundamental processes induced by the inter-layer interaction are given through
the decomposition
\begin{equation}
\lspin_{1}\cdot \lspin_{2}={1\over 2}(S^{+}_{1}S^{-}_{2}+S^{-}_{1}S^{+}_{2})+S^{z}_{1}S^{z}_{2}.
\label{eqn:inter-layerHeisenberg}\end{equation}
By rewriting this expression in a second quantized form, we get
three  scattering processes between propagating spins
on each layer.
The first term of (\ref{eqn:inter-layerHeisenberg}) produces the scattering process  
with spin flip (Type-A). 
The second term of (\ref{eqn:inter-layerHeisenberg}) produces two fundamental processes; 
scattering processes between parallel spins (Type-B) and  between anti-parallel spins (Type-C).
These processes are shown in Fig.~2.
These three scattering processes can couple to the intra-layer spin fluctuations in  
each of adjacent layers.
In Fig.~3. we show how the inter-layer exchange interaction vertices
couple to the intra-layer RPA spin fluctuations.
Thus the intra-layer spin fluctuations on each layer can couple to 
each other via these scattering vertices as shown in Fig.~4.
As is shown in Fig.~4 (a), the Type-A scattering can  couple only to the intra-layer
transverse spin fluctuations. 
Further Type-B and Type-C scattering can couple only to the intra-layer
longitudinal spin fluctuations, as depicted in Fig.~4(b),(c).
Then the enhanced inter-layer interaction for each  channel
can be given respectively by:
\begin{eqnarray}
{1\over 2}J_{\perp}^{+-}(\q;i\freq)=
\gamma^{+-}(\q;i\freq){J_{0\perp}\over 2}\gamma^{+-}(\q;i\freq)
={J_{0\perp}\over 2}\gamma(\q;i\freq)^{2},
\end{eqnarray}
for the type-A scattering and
\begin{eqnarray}
{1\over 4}J_{\perp}^{\sigma\sigma}(\q;i\freq)&=&
\gamma^{\sigma,-\sigma}(\q;i\freq){J_{0\perp}\over 4}\gamma^{-\sigma,\sigma}(\q;i\freq)
-\gamma^{\sigma,\sigma}(\q;i\freq){J_{0\perp}\over 4}\gamma^{\sigma,\sigma}(\q;i\freq)\nonumber\\
& &-\gamma^{\sigma,\sigma}(\q;i\freq){J_{0\perp}\over 4}\gamma^{-\sigma,\sigma}(\q;i\freq)
+\gamma^{\sigma,\sigma}(\q;i\freq){J_{0\perp}\over 4}\gamma^{\sigma,\sigma}(\q;i\freq)\nonumber\\
&=&{J_{0\perp}\over 4}\gamma(\q;i\freq)^{2},
\end{eqnarray}
and 
for the type-B scattering
\begin{eqnarray}
-{1\over 4}J_{\perp}^{\sigma,-\sigma}(\q;i\freq)&=&
-\gamma^{\sigma,-\sigma}(\q;i\freq){J_{0\perp}\over 4}\gamma^{\sigma,-\sigma}(\q;i\freq)
+\gamma^{\sigma,-\sigma}(\q;i\freq){J_{0\perp}\over 4}\gamma^{-\sigma,-\sigma}(\q;i\freq)\nonumber\\
&+&\gamma^{\sigma,\sigma}(\q;i\freq){J_{0\perp}\over 4}\gamma^{\sigma,-\sigma}(\q;i\freq)
-\gamma^{\sigma,\sigma}(\q;i\freq){J_{0\perp}\over 4}\gamma^{-\sigma,-\sigma}(\q;i\freq)\nonumber\\
&=&-{J_{0\perp}\over 4}\gamma(\q;i\freq)^{2},
\end{eqnarray}
for the type-C scattering.
Thus scattering vertices corresponding to each channel are
enhanced in the same manner. This situation comes from the   rotational symmetry of the interaction in the spin space.
Thus we introduce the enhanced inter-layer interaction
\begin{equation}
J_{\perp}(\q;i\freq)\equiv
J_{0\perp}\gamma(\q;i\freq)^{2}.\label{eqn:enhanced}
\end{equation}
We can see directly from these expressions that the inter-layer Heisenberg 
interaction can be strongly enhanced by the strongly enhanced intra-layer
spin fluctuations.

\section{Inter-layer Exchange Scattering Processes}
When we consider the spin fluctuating proceeses driven by the
strongly enhanced  inter-layer interaction, we should  bear in mind
that since in our treatment we neglect the inter-layer carrier hopping,
the electron-hole pair bubble laid across adjacent layers doesn't exist.
Therefore the most dominant  process driven by the inter-layer interction 
is the electron-hole exchange scattering  where an electron and a hole run 
in {\it different} layers.
This kinds of processes are expected to develop dramatically 
just near
the antiferromagnetic phase boundary and  modify the simple RPA
susceptibility.
These scattering processes are described in terms of the   T-matrices defined
by
\begin{equation}
{\cal T}^{\mu\nu,\lambda\rho}_{\perp}(k,k';q)=
\int_{0}^{1/T}d\tau e^{i\freq\tau}
<T_{\tau}[c^{\dagger}_{1,{\footnotesize\k}',\mu}(\tau)c_{1,{\footnotesize\k'+\q},\nu}(\tau)
c^{\dagger}_{2,{\footnotesize\k+\q},\rho}(0)c_{2,{\footnotesize\k},\lambda}(0)]>,\label{eqn:defofT}
\end{equation}
where $\mu,\nu,\lambda,\rho=\up$ or $\down$.

\subsection{Scattering T-matrices}
The general expression (\ref{eqn:defofT}) gives   spin dependent T-matrices, 
\begin{eqnarray}
\left\{\begin{array}{c}
{\cal{T}}^{\up\down,\up\down}_{\perp}(k,k;q)={\cal{T}}^{\down\up,\down\up}_{\perp}(k,k;q),\\\\
{\cal{T}}^{\up\up,\up\up}_{\perp}(k,k';q)={\cal{T}}^{\down\down,\down\down}_{\perp}(k,k';q),\\\\
{\cal{T}}^{\up\down,\down\up}_{\perp}(k,k';q)={\cal{T}}^{\down\up,\up\down}_{\perp}(k,k';q).
\end{array}\right.
\end{eqnarray}
To clarify the nature of these spin dependent  scattering processes, 
let us decompose them into the singlet and triplet channel.
Since in our case T-matrices must satisfy the spin rotational symmetry relation,
the scattering vertex can be written in a operator form in the spin space,
\begin{equation}
{\hat{\cal{T}}}_{\perp}={\cal{T}}_{\perp\rho}+{\cal{T}}_{\perp\sigma}\spin_{1}\cdot\spin_{2}
\end{equation}
where $\spin_{1}$ and $\spin_{2}$ denote the Pauli matrices that represent spin on
the layer-1 and layer-2 respectively. Then $\spin_{1}\cdot\spin_{2}$ has eigen values
1 and $-3$ that correspond to the triplet and singlet pair respectively.
We can decompose ${\cal{T}}_{\perp}$ into the singlet and triplet channel as
\begin{eqnarray}
\left\{ \begin{array}{c}
{\cal{T}}_{\perp}^{triplet}={\cal{T}}_{\perp\rho}+{\cal{T}}_{\perp\sigma},\\\\
{\cal{T}}_{\perp}^{singlet}={\cal{T}}_{\perp\rho}-3{\cal{T}}_{\perp\sigma}.
\end{array} \right.
\end{eqnarray}
Thus we get 
\begin{equation}
{\hat{\cal{T}}}_{\perp}={{\cal{T}}_{\perp}^{singlet}+3{\cal{T}}_{\perp}^{triplet}\over 4}
+{{\cal{T}}_{\perp}^{triplet}-{\cal{T}}_{\perp}^{singlet}\over 4}\spin_{1}\cdot\spin_{2}.
\end{equation}
On the other hand, by noting 
$\spin_{1}\cdot\spin_{2}=2(\sigma_{1}^{+}\sigma_{2}^{-}+\sigma_{1}^{-}\sigma_{2}^{+})
+\sigma_{1}^{z}\sigma_{2}^{z}$, we  get
\begin{eqnarray}
\left\{\begin{array}{c}
{\cal{T}}^{\up\down,\up\down}_{\perp}={1\over 2}\left({\cal{T}}_{\perp}^{singlet}+{\cal{T}}_{\perp}^{triplet}\right),\\\\
{\cal{T}}^{\up\up,\up\up}_{\perp}={\cal{T}}_{\perp}^{triplet},\\\\
{\cal{T}}^{\up\down,\down\up}_{\perp}={1\over 2}\left({\cal{T}}_{\perp}^{triplet}-{\cal{T}}_{\perp}^{singlet}\right).
\end{array}\right.
\end{eqnarray}
From these equations we can directly check the spin rotational invariance relation
\begin{equation}
{\cal{T}}^{\up\up,\up\up}_{\perp}-{\cal{T}}^{\up\down,\down\up}_{\perp}
={\cal{T}}^{\up\down,\up\down}_{\perp}.\end{equation}
Finally we can combine ${\cal{T}}_{\perp}^{singlet}$ and ${\cal{T}}_{\perp}^{triplet}$
with ${\cal{T}}_{\perp}^{\mu\nu,\lambda\rho}$ like
\begin{eqnarray}
{\cal{T}}_{\perp}^{singlet}&=&{\cal{T}}^{\up\down,\up\down}-{\cal{T}}^{\up\down,\down\up},\label{eqn:SingletT}\\
{\cal{T}}_{\perp}^{triplet}&=&{\cal{T}}^{\up\up,\up\up}\label{eqn:TripletT}.
\end{eqnarray}

In Fig.~5 we show the diagrammatic representation for the particel-hole ladder processes 
  where
the $n$-th order ladders are shown.
As is shown in Fig.5(a), the  type-C process
produces 
${\cal{T}}^{\up\down,\up\down}_{\perp(n)}(k,k;q)$.
As shown in  Fig.5(b), the type-A and type-B process
produce 
${\cal{T}}^{\up\up,\up\up}_{\perp(n)}(k,k';q)$ and
${\cal{T}}^{\up\down,\down\up}_{\perp(n)}(k,k';q)$.
In  ${\cal{T}}^{\up\up,\up\up}_{\perp}(q)$,
the initial spins of electron and hole are not flipped in the final state.
On the other hand in  
${\cal{T}}^{\up\down,\down\up}_{\perp}(k,k';q)$,
the initial spins  are flipped in the final state.
We note that although  ${\cal{T}}^{\up\down,\up\down}_{\perp}$
includes only one scattering channel (type-C scattering) in all the
intermediate states,
${\cal{T}}^{\up\up,\up\up}_{\perp}$  and 
${\cal{T}}^{\up\down,\down\up}_{\perp}$
include two different scattering channels (type-A and type-B scattering)
{\it in all the possible ways}.
As a result the number of  possible diagrams corersponding to Fig.6(b)
increases more and more as the order of the diagram increases.

\subsection{${\cal{T}}^{\up\down,\up\down}_{\perp}$}

We first consider ${\cal{T}}^{\up\down,\up\down}_{\perp}(k,k';q)$.
For example the 3rd order contribution can be explicitly written  by
\begin{equation}
{\cal{T}}^{\up\down,\up\down}_{\perp(3)}(k,k';q)=-{1\over 64}\sum_{k_{1},k_{2}}
J_{\perp}(k-k_{1})J_{\perp}(k_{1}-k_{2})J_{\perp}(k_{2}-k')
{\cal{G}}_{0}(k_{1}){\cal{G}}_{0}(k_{1}+q){\cal{G}}_{0}(k_{2}){\cal{G}}_{0}(k_{2}+q).
\label{eqn:3rd}\end{equation}
To proceed with our analysis, we need two approximations.
First we neglect the frequency dependence of the
interaction  and replace 
$J_{\perp}(q)$ with the most dominant contribution
$J_{\perp}({\q})\equiv J_{\perp}(\q;i\omega_{l}=0)$.
Then we can perform the matsubara summations with regard to  $k_{1}$ and $k_{2}$ in 
(\ref{eqn:3rd}) and get
\begin{eqnarray}
{\cal{T}}^{\up\down,\up\down}_{\perp(3)}(k,k';q)=-{1\over 64}\sum_{\footnotesize\k_{1},\footnotesize\k_{2}}
J_{\perp}({\k}-{\k}_{1})
\chi_{\footnotesize\k_{1}}(q;0)
J_{\perp}({\k}_{1}-{\k}_{2})
\chi_{\footnotesize\k_{2}}(q;0)
J_{\perp}({\k}_{2}-{\k}').
\end{eqnarray}
Now we introduced the quantity $\chi_{\k}(q;Q)$ by
\begin{equation}
\chi_{\footnotesize\k}(q;Q)\equiv- T\sum_{\varepsilon_{n}}{\cal G}_{0}(k+q){\cal G}_{0}(k+Q)=
-{1\over 2}{\tanh{\xi_{\footnotesize\k+\q}\over 2T}-\tanh{\xi_{\footnotesize\k+\Q}\over 2T}
\over i\omega_{l}-i\Omega_{m}-\eta_{\footnotesize\k}(\q;\Q)}.\label{eqn:chiqQ}
\end{equation}
We note that $J_{\perp}(\q)$ has a broad maximum around the antiferromagnetic vector
$\q^{*}=(\pi/q,\pi/a)$.
Furthermore the internal wavenumber ${\k}$ and  ${\k}'$ become important 
only when they satisfy the condition
 ${\k}-{\k}'\sim \bf{0}$ ({\it forward} scattering) or 
 ${\k}- {\k}'\sim \q^{*}$ ({\it backward} scattering).
The  latter  process  is strongly driven by the antiferromagnetic spin fluctuations. 
We can see that the 3rd order process is enhanced only in the case of 
backward scattering,
since    ${\k}- {\k}_{1}\sim \q^{*}$ and 
${\k}_{1}- {\k}_{2}\sim \q^{*}$ give constraint on ${\k}'$ as
${\k}'\sim {\k}+\q^{*}$.

This consideration gives us important results:
the ladders with  odd number $J_{\perp}(\q)$  give dominant contribution
to  backward scattering processes, while 
the ladders with  even number $J_{\perp}(\q)$ 
 give dominant contribution to  forward scattering processes.
This situations are shown in Fig.~6
Keeping this constraints in mind, we 
replace $J_{\perp}(\q)$ with the averaged one
\footnote{To replace the non-separable momentum dependent interaction with 
its averaged value
is inevitable to solve the Bethe-Salpeter equation in a closed form. 
In the case of Coulomb gas,
 the dielectric screening removes the singularity and
as a result this approximation may work well. Wolff discussed the spin susceptibility
of electron gas in this manner\cite{Wolff}.

In the present problem, the spin-fluctuation mediated interaction
$J_{\perp}(q)$ has short range nature and doesn't have any singularity
in a paramagnetic phase. Therefore in the present case this approximation
can be justified. 
}
\begin{eqnarray}
{\cal J}_{\perp}(T)=\sum_{\q} J_{\perp}(\q, i\omega_{l}=0)
=J_{0\perp}\sum_{\q}\{\gamma(\q, i\omega_{l}=0)\}^{2}
\label{eqn:average},
\end{eqnarray}
which depends  on the temperature through $\chi_{0}(\q;0)$ in the Stoner factors.
This procedure is our second approximation.
Now ${\cal{T}}^{\up\down,\up\down}_{\perp(3)}(k,k';q)$ can be approximated simply by
\begin{eqnarray}
{\cal{T}}^{\up\down,\up\down}_{\perp(3)}(k,k';q)=-\left\{ {{\cal J}_{\perp}(T) \over 4}\right\}^{3}
\chi_{0}(q)^{2},
\end{eqnarray}
where $\chi_{0}(q)=\sum_{\k}\chi_{\k}(q,0)$.
At this stage T-matrix depends only on $q$. Thus from now on we simply denote  the T-matrix
as  ${\cal{T}}^{\up\down,\up\down}_{\perp}(q)$. 

Genelarizing the aformentioned procedure, we can get the T-matrixes for forward and backward process as
\begin{eqnarray}
{\cal{T}}^{\up\down,\up\down}_{\perp{{FW}}}(q)=
{\left\{ { {\cal J}_{\perp}(T)\over 4 }\right\}^{2}\chi_{0}(q)
\over 1-\left\{ {{\cal J}_{\perp}(T)\over 4}\chi_{0}(q)\right\}^{2} }
\label{eqn:gammaev},
\end{eqnarray}
\begin{eqnarray}
{\cal{T}}^{\up\down,\up\down}_{\perp{{BW}}}(q)=
-{ \left\{{{\cal J}_{\perp}(T)\over 4}\right\}^{3} \chi_{0}(q)^{2}
\over 1-\left\{ {{\cal J}_{\perp}(T)\over 4}\chi_{0}(q)\right\}^{2}}\label{eqn:gammaod},
\end{eqnarray}
where FW and BW denote the case of forward and backward process respectively.

\subsection{${\cal{T}}^{\up\up,\up\up}_{\perp}$ and ${\cal{T}}^{\up\down,\down\up}_{\perp}$}

Next we consider
${\cal{T}}^{\up\up,\up\up}_{\perp(n)}$ and ${\cal{T}}^{\up\down,\down\up}_{\perp(n)}$.
To get explicit expressions	 for them, let us introduce the auxiliary spin-independent T-matrix
${\cal{T}}_{\perp(n)}$ that is defined  in Fig.5(c).
In  ${\cal{T}}_{\perp(n)}$,  all the inter-plane interactions
contained in the ladder are type-B interaction.
In reality ${\cal{T}}^{\up\up,\up\up}_{\perp(n)}$ and ${\cal{T}}^{\up\down,\down\up}_{\perp(n)}$
contain type-A and type-B interactions {\it in all the possible ways}.
Then  $n$-th order ladder ${\cal{T}}^{\up\up,\up\up}_{\perp(n)}$
 and ${\cal{T}}^{\up\down,\down\up}_{\perp(n)}$
can simply be related to corresponding ${\cal{T}}_{\perp(n)}$
through combinatrics factor as 
\begin{eqnarray}
{\cal{T}}^{\up\up,\up\up}_{\perp(n)}&=&a_{n}
{\cal{T}}_{\perp(n)}\label{eqn:an}\\
{\cal{T}}^{\up\down,\down\up}_{\perp(n)}&=&b_{n}
{\cal{T}}_{\perp(n)}
\end{eqnarray}
where
\begin{eqnarray}
a_{n}&=&\sum_{i=0}^{[{n\over 2}]}2^{2i}{_{n}}C_{2i},\label{eqn:exa}\\
b_{n}&=&\sum_{i=0}^{[{n\over 2}-{1\over 2}]}2^{2i+1}{_{n}}C_{2i+1}\label{eqn:exb}.
\end{eqnarray}
Here $_{n}C_{m}={n!/[(m-n)!m!]}$ is a binomial coefficient.

The above relations have already been discussed in Ref.\cite{kishine}. Now, for example,  we derive the formula (\ref{eqn:an}).
In this case $n$-th order term of ${\cal{T}}^{\up\up,\up\up}_{\perp(n)}$ 
can include 
the {\it even} number spin-flipping (Type-A)  scattering. Then all the other
scatterings are  the type-B scattering.
We consider the case when there are $2i\,\,(i=0,1,\cdots)$ type-A processses. Then there are
$n-2i$ Type-B vertices in the $n$-th order ladder.
Since the type-A vertex gives the facor $J_{0\perp}/2$ and  
the type-B vertex gives the facor $J_{0\perp}/4$, if we replace  all the
type-A vertices simply with the Type-B vertices to get  ${\cal{T}}_{\perp(n)}$, 
there appear the factor
$2^{2i}$. 
Furthermore, concerning the location of the type-A
interactions in a ladder there are $_{n}C_{2i}$ ways.
As a result 
${\cal{T}}^{\up\up,\up\up}_{\perp(n)}$ is related with ${\cal{T}}_{\perp(n)}$ through a factor
$\sum_{i=0}^{[{n\over 2}]}2^{2i}{_{n}}C_{2i}$.

Now we notice that $n$-th order prefactors satisfy the
simple relations
\begin{equation}
a_{n}+b_{n}=(1+2)^{n}=3^{n},\label{eqn:combi}
\end{equation}
and
\begin{equation}
a_{n}-b_{n}=(1-2)^{n}=(-1)^{n}\label{eqn:srsg}.
\end{equation}
We can get actual feeling for these relations by writing down
concrete values of multiplication factors
\begin{eqnarray*}
\left\{
\begin{array}{l}
a_{2}=5\\
b_{2}=4
\end{array}\right.,
\left\{
\begin{array}{l}
a_{3}=13\\
b_{3}=14
\end{array}\right.,
\left\{
\begin{array}{l}
a_{4}=41\\
b_{4}=40
\end{array}\right.,
\left\{
\begin{array}{l}
a_{5}=121\\
b_{5}=122
\end{array}\right., \cdots.
\end{eqnarray*}
Using the relation, (\ref{eqn:combi}), we can obtain
\begin{equation}
{\cal{T}}^{\up\up,\up\up}_{\perp(n)}+{\cal{T}}^{\up\down,\down\up}_{\perp(n)}
=3^{n}
{\cal{T}}_{\perp(n)}\label{GammaUUUD}.
\end{equation}
Therefore we can get the explicit result for ${\cal{T}}^{\up\up,\up\up}_{\perp}+{\cal{T}}^{\up\down,\down\up}_{\perp}$.
In the case of forward process,  dominant contributions arise from
\begin{eqnarray}
{\cal{T}}^{\up\up,\up\up}_{\perp FW}+{\cal{T}}^{\up\down,\down\up}_{\perp FW}=
{\left\{{3\over 4}{{\cal J}}_{\perp}(T)\right\}^{2}\chi_{0}(q)\over 1-
\left\{{3\over 4}{{\cal J}}_{\perp}(T)\chi_{0}(q)\right\}^{2}},\label{eqn:sume}
\end{eqnarray}
while in the case of backward process, dominant contributions arise from
\begin{eqnarray}
{\cal{T}}^{\up\up,\up\up}_{\perp BW}+{\cal{T}}^{\up\down,\down\up}_{\perp BW}=
{\left\{{3\over 4}{{\cal J}}_{\perp}(T)\right\}^{3}\chi_{0}(q)^{2}\over 1-
\left\{{3\over 4}{{\cal J}}_{\perp}(T)\chi_{0}(q)\right\}^{2}}\label{eqn:sumo}.
\end{eqnarray}

Next we go back to the relation (\ref{eqn:srsg}) that gives
 the simple relation
\begin{equation}
{\cal{T}}^{\up\up,\up\up}_{\perp(n)}-{\cal{T}}^{\up\down,\down\up}_{\perp(n)}=(-1)^{n}
{\cal{T}}_{\perp(n)}.
\end{equation}
Now we can notice that
\begin{equation}
(-1)^{n}
{\cal{T}}_{\perp(n)}=
{\cal{T}}^{\up\down,\up\down}_{\perp(n)}.
\end{equation}
Therefore the relation (\ref{eqn:srsg}) directly leads to the spin-rotational symmetry relation
\begin{equation}
{\cal{T}}^{\up\up,\up\up}_{\perp}
-{\cal{T}}^{\up\down,\down\up}_{\perp}
={\cal{T}}^{\up\down,\up\down}_{\perp}.\label{eqn:SRS2}
\end{equation}
Thus we can confirm {\it again} that the present treatment
 doesn't violate the spin rotaional symmetry.

The relations (\ref{eqn:sume}), (\ref{eqn:sumo}) and (\ref{eqn:SRS2}) enable us to get the explicit form of 
${\cal{T}}^{\up\up,\up\up}_{\perp}$ and ${\cal{T}}^{\up\down,\down\up}_{\perp}$. 
We get
\begin{eqnarray}
{\cal{T}}^{\up\up,\up\up}_{\perp FW}(q)=
{1\over 2}\left[
{ \left\{ {3\over4}{{\cal J}}_{\perp}(T)\right\}^{2}\chi_{0}(q)\over
 1-\left\{{3\over4}{{\cal J}}_{\perp}(T)\chi_{0}(q)\right\}^{2}}
+
{\left\{ {1\over4}{{\cal J}}_{\perp}(T)\right\}^{2}\chi_{0}(q)\over
 1-\left\{{1\over4}{{\cal J}}_{\perp}(T)\chi_{0}(q)\right\}^{2}}
\right],\label{FW1}\\
{\cal{T}}^{\up\down,\down\up}_{\perp FW}(q)=
{1\over 2}\left[
{\left\{ {3\over4}{{\cal J}}_{\perp}(T)\right\}^{2}\chi_{0}(q)\over
 1-\left\{{3\over4}{{\cal J}}_{\perp}(T)\chi_{0}(q)\right\}^{2}}
-
{\left\{ {1\over4}{{\cal J}}_{\perp}(T)\right\}^{2}\chi_{0}(q)\over
 1-\left\{{1\over4}{{\cal J}}_{\perp}(T)\chi_{0}(q)\right\}^{2}}
\right],
\end{eqnarray}
for forward process and
\begin{eqnarray}
{\cal{T}}^{\up\up,\up\up}_{\perp BW}(q)=
{1\over 2}\left[
{\left\{{3\over4}{{\cal J}}_{\perp}(T)\right\}^{3}\chi_{0}(q)^{2}\over
 1-\left\{{3\over4}{{\cal J}}_{\perp}(T)\chi_{0}(q)\right\}^{2}}
-
{\left\{{{1\over4}{\cal J}}_{\perp}(T)\right\}^{3}\chi_{0}(q)^{2}\over
 1-\left\{{1\over4}{{\cal J}}_{\perp}(T)\chi_{0}(q)\right\}^{2}}
\right],\\
{\cal{T}}^{\up\down,\down\up}_{\perp BW}(q)=
{1\over 2}\left[
{\left\{{3\over4}{{\cal J}}_{\perp}(T)\right\}^{3}\chi_{0}(q)^{2}\over
 1-\left\{{3\over4}{{\cal J}}_{\perp}(T)\chi_{0}(q)\right\}^{2}}
+
{\left\{{{1\over4}{\cal J}}_{\perp}(T)\right\}^{3}\chi_{0}(q)^{2}\over
 1-\left\{{1\over4}{{\cal J}}_{\perp}(T)\chi_{0}(q)\right\}^{2}}
\right]\label{BW2},
\end{eqnarray}
for backward process.

\subsection{${\cal{T}}^{singlet}_{\perp}$ and ${\cal{T}}^{triplet}_{\perp}$}

By using (\ref{eqn:SingletT}), (\ref{eqn:TripletT}) and (\ref{FW1}) $\sim$ (\ref{BW2}), we can get the explicit form of the T-matrices in the singlet and triplet channel
in the case of the forward and backward process as
\begin{eqnarray}
{\cal{T}}^{singlet}_{\perp FW}(q)=
{3\over 2}
{ \left\{ {1\over4}{{\cal J}}_{\perp}(T)\right\}^{2}\chi_{0}(q)\over
 1-\left\{{1\over4}{{\cal J}}_{\perp}(T)\chi_{0}(q)\right\}^{2}}
-{1\over2}
{\left\{ {3\over4}{{\cal J}}_{\perp}(T)\right\}^{2}\chi_{0}(q)\over
 1-\left\{{3\over4}{{\cal J}}_{\perp}(T)\chi_{0}(q)\right\}^{2}},\\
{\cal{T}}^{triplet}_{\perp FW}(q)=
{1\over 2}
{ \left\{ {3\over4}{{\cal J}}_{\perp}(T)\right\}^{2}\chi_{0}(q)\over
 1-\left\{{3\over4}{{\cal J}}_{\perp}(T)\chi_{0}(q)\right\}^{2}}
+{1\over2}
{\left\{ {1\over4}{{\cal J}}_{\perp}(T)\right\}^{2}\chi_{0}(q)\over
 1-\left\{{1\over4}{{\cal J}}_{\perp}(T)\chi_{0}(q)\right\}^{2}},
\end{eqnarray}
and
\begin{eqnarray}
{\cal{T}}^{singlet}_{\perp BW}(q)=
-{3\over 2}
{ \left\{ {1\over4}{{\cal J}}_{\perp}(T)\right\}^{3}\chi_{0}(q)^{2}\over
 1-\left\{{1\over4}{{\cal J}}_{\perp}(T)\chi_{0}(q)\right\}^{2}}
-{1\over2}
{\left\{ {3\over4}{{\cal J}}_{\perp}(T)\right\}^{3}\chi_{0}(q)^{2}\over
 1-\left\{{3\over4}{{\cal J}}_{\perp}(T)\chi_{0}(q)\right\}^{2}},\\
{\cal{T}}^{triplet}_{\perp BW}(q)=
{1\over 2}
{ \left\{ {3\over4}{{\cal J}}_{\perp}(T)\right\}^{3}\chi_{0}(q)^{2}\over
 1-\left\{{3\over4}{{\cal J}}_{\perp}(T)\chi_{0}(q)\right\}^{2}}
-{1\over2}
{\left\{ {1\over4}{{\cal J}}_{\perp}(T)\right\}^{3}\chi_{0}(q)^{2}\over
 1-\left\{{1\over4}{{\cal J}}_{\perp}(T)\chi_{0}(q)\right\}^{2}}.
\end{eqnarray}

In Fig.~7, we show the behavior of the singlet and triplet T-matrices
as a function of the dimensionless quantity, ${\cal J}_{\perp}(T)\chi_{0}(q)/4$.
It follows that the inter-layer  exchange scattering processes 
give rise to the strong attraction for the spin singlet pair and
the strong repulsion for the spin triplet pair.
Thus we can say that {\it just near the intra-layer antiferromagnetic phase boundary
the inter-layer singlet pair formation is strongly favored while the triplet pair
formation is strongly prohibited.}
This tendency mainly comes from the dramatically enhanced repulsion between the inter-layer parallel spin pair,
${\cal{T}}^{\up\up,\up\up}_{\perp}(q)$ and  ${\cal{T}}^{\up\down,\down\up}_{\perp}(q)$.
We can expect that the strong repulsion for inter-layer parallel spin pairs induces the screening
effects for the intra- and inter-layer spin fluctuations. We investigate this
screening effects in the next section.

\subsection{Averaged Inter-layer Interaction ${\cal J}_{\perp}(T)$}

It is should be noted that the Stoner factor sensitively depends on the magnitude
of the intra-layer Hubbard interaction.
Our proposal can be justified only when the Stoner factor becomes considerablly
large.
In Fig.~8. we show the numerical results for the averaged inter-layer interaction
${\cal{J}}_{\perp}$ for $U=3.25 t$.
Here the half-filling corresponds to $\mu=-0.327 t$.
We fixed the chemical potential to $\mu=-0.54 t$.
By taking up the avarage, we can also check the non-divergence of the Stoner factor
over the whole Brillouin zone.  
If the bare susceptibility touches the critical condition, $U\chi_{0}(q)=1$, at a certain temperature, $T_{\rm SDW}$, 
the RPA susceptibility diverges and the whole theoretical backgrounds break down.

As for the magnitude of $J_{0\perp}$, we set $J_{0\perp}=0.08t$.
This magnitude of the bare inter-layre Heisenberg interaction corresponds
to the result suggested in Ref.\cite{midinfra}.
The maximam of the bare susceptibility, $\chi_{0}(q)$,
reach $0.26/t$. Therefore, the divergence of  ${\cal{T}}^{s inglet, triplet}_{\perp}$ occurs at
$0.75{\cal{J}}_{\perp}\chi_{0}=1$ that corresponds to $0.75{\cal J}_{\perp}\sim 3.8$.
As we can see from the result of Fig.~8, the numerical vale of $0.75{\cal J}_{\perp}$
can remain finite down to $T=0$ with the maximum value, $0.75{\cal J}_{\perp}\sim 3.9$. 
If we take larger magnitude of $U$,  ${\cal J}_{\perp}$  becomes divergent
 at unreallistically high temperature, $T_{SDW}$.

\section{Contributions of Exchange Scattering 
Processes to Diagonal and Off-diagonal Susceptibility}

Now we consider explicitly 
how inter-layer exchange scattering processees affect the irreducible diagonal- and
off-diagonal susceptibilities.
Within the present scheme,  the inter-layer exchange processes
cannot be detected {\it directly} through experimental probe
because in our model there is no inter-layer direct carrier hopping.
Instead these processes are inserted into the irreducible polarization. 
In Fig.~9 we show 
the  contributions from exchange scattering T-matrices to the transverse
susceptibility up to the 2nd order.
Among them (a),(b), (c)  and (d) contribute to the diagonal counterpart of 
the transverse 
susceptibility, $\Delta\chi^{+-}_{//}$, although (e),(f), and (g) contribute to the
off-diagonal counterpart, $\Delta\chi^{+-}_{\perp}$.
In the appendix  we shall discuss
the longitudinal counterpart, $\Delta\chi^{zz}_{//}$ and $\Delta\chi^{zz}_{\perp}$,
and prove the spin rotational invariance relation.
As we shall discuss in the next section, 			
we can notice that the neutron scattering experiments
detect all diagrams in Fig.~9, while the NMR longitudinal relaxation experiments detect
only diagram (a),(b),(c), and (d).
Furthermore all diagrams  contribute to the NMR spin echo decay rate.
We note that in each diagram the forward and backward T-matrices are inserted correspondong to
the characteristic momentum transfer.
Thus we get the expressions for the correction terms to the diagonal counterpart as
\begin{eqnarray}
\Delta\chi^{+-}_{//}=
\Delta\chi^{+-}_{//,\rm a}+\Delta\chi^{+-}_{//,\rm b}+
\Delta\chi^{+-}_{//,\rm c}+\Delta\chi^{+-}_{//,\rm d},\label{eqn:correctionparallel}
\end{eqnarray}
where
\begin{eqnarray*}
\Delta\chi^{+-}_{//,\rm a}&=&
-\gamma(q)^{2}T
\displaystyle\sum_{\Omega_{m}}\sum_{\footnotesize\Q}{\cal{T}}^{\up\up,\up\up}_{\perp FW}(Q)
{\cal L}^{(1)}(q,Q),\\
\Delta\chi^{+-}_{//,\rm b}&=&-\gamma(q)^{2}T
\displaystyle\sum_{\Omega_{m}}\sum_{\footnotesize\Q}
{\cal{T}}^{\up\down,\up\down}_{\perp FW}(Q)
{\cal L}^{(1)}(q,Q),\\
\Delta\chi^{+-}_{//,\rm c}&=&\gamma(q)^{2}T
\displaystyle\sum_{\Omega_{m}}\sum_{\footnotesize\Q}
\left\{
{\cal{T}}^{\up\up,\up\up}_{\perp FW}(Q){\cal{T}}^{\up\down,\up\down}_{\perp FW}(Q+q)+
{\cal{T}}^{\up\up,\up\up}_{\perp BW}(Q){\cal{T}}^{\up\down,\up\down}_{\perp BW}(Q+q)\right\}
\nonumber\\
&&\,\,\,\,\,\,\,\,\,\,\,\,\,\,\,\,\,\,\,\,\,\,\,\,\,\,\,\,\,\,\,\,\,\,\,\,\,\,\,\,\,\,\,\,\,\,\,\,\,\,\,\,\,\,\,\,\,\,\,\,\,\,\,\,\,\,\,\,\,\,
\,\,\,\,\,\,\,\,\,\,\,\,\,\,\,\,\,\,\,\,\,\,\,\,\,\,\,\,\,\,\,\,\,\,\,\,\,\,\,\,\,\,
\times\{{\cal L}^{(3)}(q,Q)\}^{2},\\
\Delta\chi^{+-}_{//,\rm d}&=&\gamma(q)^{2}T
\displaystyle\sum_{\Omega_{m}}\sum_{\footnotesize\Q}
\left\{
{\cal{T}}^{\up\down,\down\up}_{\perp FW}(Q){\cal{T}}^{\down\down,\down\down}_{\perp FW}(Q+q)+
{\cal{T}}^{\up\down,\down\up}_{\perp BW}(Q){\cal{T}}^{\down\down,\down\down}_{\perp BW}(Q+q)\right\}
\nonumber\\
&&\,\,\,\,\,\,\,\,\,\,\,\,\,\,\,\,\,\,\,\,\,\,\,\,\,\,\,\,\,\,\,\,\,\,\,\,\,\,\,\,\,\,\,\,\,\,\,\,\,\,\,\,\,\,\,\,\,\,\,\,\,\,\,\,\,\,\,\,\,\,
\,\,\,\,\,\,\,\,\,\,\,\,\,\,\,\,\,\,\,\,\,\,\,\,\,\,\,\,\,\,\,\,\,\,\,\,\,\,\,\,\,
\times\{{\cal L}^{(3)}(q,Q)\}^{2}.
\end{eqnarray*}
Each indices a, b, c, ... represent the corresponding diagrams in Fig.~9.
Furthermore the correction terms to the off-diagonal counterpart are given by
\begin{eqnarray}
\Delta\chi^{+-}_{\perp}=
\Delta\chi^{+-}_{\perp,\rm e}+\Delta\chi^{+-}_{\perp,\rm f}+
\Delta\chi^{+-}_{\perp,\rm g},\label{eqn:correctionperp}
\end{eqnarray}
where
\begin{eqnarray*}
\Delta\chi^{+-}_{\perp,\rm e}&=&-\gamma(q)^{2}
T\displaystyle\sum_{\Omega_{m}}\sum_{\footnotesize\Q}
{\cal{T}}^{\up\down,\down\up}_{\perp BW}(Q)
{\cal L}^{(2)}(q,Q),\\
\Delta\chi^{+-}_{\perp,\rm f}&=&\gamma(q)^{2}T
\displaystyle\sum_{\Omega_{m}}\sum_{\footnotesize\Q}
[
{\cal{T}}^{\up\down,\up\down}_{\perp FW}(Q){\cal{T}}^{\up\down,\down\up}_{\perp BW}(Q+q)
+{\cal{T}}^{\up\down,\up\down}_{\perp BW}(Q){\cal{T}}^{\up\down,\down\up}_{\perp FW}(Q+q)\\
+&&\!\!\!\!\!\!\!\!\!\!\!\!
{\cal{T}}^{\up\down,\up\down}_{\perp FW}(Q){\cal{T}}^{\up\down,\down\up}_{\perp BW}(Q+q)
+{\cal{T}}^{\up\down,\up\down}_{\perp BW}(Q){\cal{T}}^{\up\down,\down\up}_{\perp FW}(Q+q)
]
{\cal L}^{(3)}(q,Q){\cal L}^{(3)}(-q,-Q),\nonumber\\
\Delta\chi^{+-}_{\perp,\rm g}&=&\gamma(q)^{2}T
\displaystyle\sum_{\Omega_{m}}\sum_{\footnotesize\Q}
[
{\cal{T}}^{\up\down,\down\up}_{\perp FW}(Q){\cal{T}}^{\up\down,\up\down}_{\perp BW}(Q+q)
+{\cal{T}}^{\up\down,\down\up}_{\perp BW}(Q){\cal{T}}^{\up\down,\up\down}_{\perp FW}(Q+q)\\
+&&\!\!\!\!\!\!\!\!\!\!\!\!
{\cal{T}}^{\up\down,\down\up}_{\perp FW}(Q){\cal{T}}^{\up\down,\up\down}_{\perp BW}(Q+q)
+{\cal{T}}^{\up\down,\down\up}_{\perp BW}(Q){\cal{T}}^{\up\down,\up\down}_{\perp FW}(Q+q)
]{\cal L}^{(3)}(q,Q){\cal L}^{(3)}(-q,-Q).\nonumber
\end{eqnarray*}
We note that in the diagram (c), (d), (f), and (g),   
forward and backward counterparts are combined to each other. 
The coupling functions are given by
\begin{eqnarray}
{\cal L}^{(1)}(\q;i\omega_{l},\Q;i\Omega_{m})={T}
\sum_{\varepsilon_{n}}\sum_{\footnotesize\k}
\{{\cal G}_{0}(k)\}^{2}{\cal G}_{0}(k+q){\cal G}_{0}(k+Q)\nonumber\\
={T}\sum_{\varepsilon_{n}}\sum_{\footnotesize\k}
{{\cal G}_{0}(k)-{\cal G}_{0}(k+q)\over i\freq-\xi_{\footnotesize\k+\q}+\xi_{\footnotesize\k}}
{{\cal G}_{0}(k)-{\cal G}_{0}(k+Q)\over i\Omega_{m}-\xi_{\footnotesize\k+\Q}+\xi_{\footnotesize\k}}\nonumber\\
=\sum_{\footnotesize\k}
{-{1\over 4T}\cosh^{-2}{\xi_{\footnotesize\k}\over2T}
+\chi_{\footnotesize\k}(0;Q)+\chi_{\footnotesize\k}(q;0)-\chi_{\footnotesize\k}(q;Q)\over 
[i\freq-\xi_{\footnotesize\k+\footnotesize\q}+\xi_{\footnotesize\k}]
 [i\Omega_{m}-\xi_{\footnotesize\k+\footnotesize\Q}+\xi_{\footnotesize\k}]},
\end{eqnarray}
\begin{eqnarray}
{\cal L}^{(2)}(\q;i\omega_{l},\Q;i\Omega_{m})={T}\sum_{\varepsilon_{n}}\sum_{\footnotesize\k}
{\cal G}_{0}(k){\cal G}_{0}(k+q){\cal G}_{0}(k+Q){\cal G}_{0}(k+Q+q)\nonumber\\
={T\over N}\sum_{\varepsilon_{n}}\sum_{\footnotesize\k}
{{\cal G}_{0}(k)-{\cal G}_{0}(k+q) \over i\omega_{l}-\eta_{\footnotesize\k}(\q,0)}
{{\cal G}_{0}(k+Q)-{\cal G}_{0}(k+q+Q) \over i\omega_{l}-\eta_{\footnotesize\k}({\footnotesize\q+\Q,\Q})}\nonumber\\
=\sum_{\footnotesize\k}
{-\chi_{\k}(0;Q)+\chi_{\footnotesize\k}(0;q+Q)+\chi_{\footnotesize\k}(q;Q)-\chi_{\footnotesize\k}(q;q+Q)
\over [i\omega_{l}-\eta_{\footnotesize\k}({\footnotesize\q},0)]  [i\omega_{l}-
\eta_{\footnotesize\k}({\footnotesize\q}+{\footnotesize\Q,\Q})]},
\end{eqnarray}
and
\begin{eqnarray}
{\cal L}^{(3)}(\q;i\omega_{l},\Q;i\Omega_{m})=T\sum_{\varepsilon_{n}}\sum_{\footnotesize\k}
{\cal G}_{0}(k){\cal G}_{0}(k+Q){\cal G}(k+Q+q)\nonumber\\
=T\sum_{\varepsilon_{n}}\sum_{\footnotesize\k}
{{\cal G}_{0}(k){\cal G}_{0}(k+Q+q)-{\cal G}_{0}(k+Q){\cal G}_{0}(k+Q+q)\over
 i\Omega_{m}-\eta_{\footnotesize\k}(\footnotesize\Q,0)}
\nonumber\\
=\sum_{\footnotesize\k}
{-\chi_{\footnotesize\k}(0;q+Q)+\chi_{\footnotesize\k}(Q;q+Q)
\over i\Omega_{m}-\eta_{\footnotesize\k}(Q,0)},
\end{eqnarray}
where 
$
\chi_{\footnotesize\k}(q;Q)
$
has already been defined by (\ref{eqn:chiqQ}) and
$\eta_{\footnotesize\k}({\footnotesize\q,\Q})
=\xi_{\footnotesize\k+\q}-\xi_{\footnotesize\k+\Q}$.
In the above expressions,  $q=(\q,i\omega_{l})$, $Q=(\Q,i\Omega_{m})$ 
where $\omega_{l}$ and $\Omega_{m}$ denote bosonic frequencies.

In the terminology of superconducting fluctuation,
(a), (b), and (e) correspond to Maki-Thompson diagram\cite{MT}, although
(c), (d), (f), and (g) correspond to Aslamazov-Larkin diagram\cite{AL}.
 Maki-Thompson terms  represent the coupling of spin fluctuations via the
exchange scattering processes with the internal momentum transfer $\Q$, 
although Aslamazov-Larkin term
represent the  coupling  of the fluctuations via the processes
with the internal momentum transfer $\Q$ and $\Q+\q$.
Therefore  in the case of the antiferromagnetic spin fluctuations, 
at the commensurate momentum $\q=\q^{*}=(\pi/a,\pi/a)$,
the coupling functions in the Aslamazov-Larkin terms are severely reduced.
Moreover in our case  Aslamazov-Larkin terms 
inevitablly include 
 ${\cal{T}}^{\up\down,\up\down}_{\perp}$ 
that is only weakly enhanced, and therefore can safely be 
discarded.

Based on the above consideration, we can say that the dominant 
contribution to the diagonal susceptibility arises from the diagram Fig.~9(a), i.e.
\begin{eqnarray}
\Delta\chi^{+-}_{//}\sim\Delta\chi^{+-}_{//,\rm a}\label{eqn:a}.
\end{eqnarray}
Furthermore the most dominant 
contribution to the off-diagonal  counterpart arises form the diagram Fig.~9(e), i.e.
\begin{eqnarray}
\Delta\chi^{+-}_{\perp}\sim\Delta\chi^{+-}_{\perp,\rm e}.\label{eqn:e}
\end{eqnarray}
Since the dominant contribution of T-matrices, ${\cal{T}}^{\mu\nu,\lambda\rho}_{\perp FW,BW}(Q)$, 
arises from the terms
with its external frequency $\Omega_{m}=0$, from now on we put $\Omega_{m}=0$ in $Q$.
Moreover the dominant contribution of ${\cal{T}}^{\mu\nu,\lambda\rho}_{\perp FW,BW}
(\Omega_{m}=0,\Q)$ in 
(\ref{eqn:a}) and (\ref{eqn:e}) arises form the term with $\q\sim\Q\sim\q^{*}$.
Thus
\begin{eqnarray}
\Delta \chi^{+-}_{//} (\q^{*};i\omega_{l}) \sim 
-T\gamma(\q^{*};i\omega_{l})^{2}
{\cal{T}}^{\up\up,\up\up}_{\perp FW}(\q^{*};i\Omega_{m}=0)
{\cal L}^{(1)}(\q^{*};i\omega_{l};\q^{*};i\Omega_{m}=0),\label{eqn:D}
\end{eqnarray}
and
\begin{eqnarray}
\Delta \chi^{+-}_{\perp}(\q         ;i\omega_{l})\sim 
-T\gamma(\q;i\omega_{l})^{2}
{\cal{T}}^{\up\down,\down\up}_{\perp BW}(\q^{*};i\Omega_{m}=0)
{\cal L}^{(2)}(\q^{*};i\Omega_{l};\q^{*};i\Omega_{m}=0).\label{eqn:OD}
\end{eqnarray}
Then we get the explicit expression for coupling functions as
\begin{eqnarray}
{\cal L}^{(1)}(\q^{*};i\omega_{l};\q^{*};i\Omega_{m}=0)&=&
\displaystyle\sum_{\footnotesize\k}
\left[{ -{1\over 2T}\cosh^{-2}{\xi_{\footnotesize\k}\over2T} \over \freq^{2}+\eta_{\footnotesize\k}(\q^{*};0)^{2}}
+2{\eta_{\k}(\q^{*};0)\Lambda_{\footnotesize\k}(\q^{*};0)\over
 \{\freq^{2}+\eta_{\footnotesize\k}(\q^{*};0)^{2}\}^{2} }\right],\\
{\cal L}^{(2)}(\q^{*};i\omega_{l};\q^{*};i\Omega_{m}=0)&=&
\displaystyle{1\over 2}\sum_{\footnotesize\k}{1\over \eta_{\footnotesize\k}(\q^{*};0)}
{\Lambda_{\k}(\q^{*};0)\over \omega_{l}^{2}+\eta_{\k}(\q^{*},0)^{2}},\label{eqn:L2}\\
{\cal L}^{(3)}(\q^{*};i\omega_{l};\q^{*};i\Omega_{m}=0)&=&
-
\displaystyle{1\over 2}\sum_{\k}
{\Lambda_{\k}(\q^{*};0)\over \omega_{l}^{2}+\eta_{\k}(\q^{*},0)^{2}}.
\end{eqnarray}
In this case
${\cal L}^{(3)}(\q^{*};i\omega_{l};\q^{*};i\Omega_{m}=0)=0$
and thus the diagram (c), (d), (f) and (g) give no contribution.
Although we can safely say that $\Delta \chi^{+-}_{\perp}$ gives rise to
 the negative 
contribution, the sign of the contribution of $\Delta \chi^{+-}_{//}$ is ambiguous.
To settle this problem, it is rather instructive to investigate
explicitly the behavior of the integrand
of ${\cal L}^{(2)}(\q^{*};i\omega_{l};\q^{*};i\Omega_{m}=0)$, i.e.
$${\cal L}^{(1)}_{\footnotesize\k}(\q^{*};i\omega_{l};\q^{*};i\Omega_{m}=0)=
{ -{1\over 2T}\cosh^{-2}{\xi_{\footnotesize\k}\over2T} \over \freq^{2}+\eta_{\footnotesize\k}(\q^{*};0)^{2}}
+2{\eta_{\k}(\q^{*};0)\Lambda_{\footnotesize\k}(\q^{*};0)\over
 \{\freq^{2}+\eta_{\footnotesize\k}(\q^{*};0)^{2}\}^{2} }.
$$
As shown in Fig.10,
the integrand is positive in almost all the region in the first Brillouin zone.
We can see that this behavior sensitively reflects  the present Fermi surface
given by (\ref{eqn:band}).
Thus in the present case both of the diagram (a) and (e) give negative contribution
 and  suppresses the total spin fluctuations, $\chi+\Delta\chi{//}+\Delta\chi{\perp}$, in the low energy region.
We can call this mechanism the {\it low energy dynamical screening effects}.
From this  result, we can see that the coupling of the  spin fluctuations on each layer
 via the enhanced inter-layer exchange scattering processes
can dramatically {\it screen} the low energy intra-layer spin fluctuations, $\chi(q)$.

\section{Neutron Scattering and NMR in Magnetically Coupled Bi-layer Cuprates}
In the present section we show how the diagonal and off-diagonal 
spin-spin correlation functions
appear in the expression of the neutron scattering 
cross section and NMR relaxation rates.

\begin{flushleft}
{\underline{\it Neutron Scattering Cross Section}}
\end{flushleft}

A neutron has magnetic moment 
${\mbox{\boldmath$\mu$}}_{N}=-\gamma \mu_{N}{\mbox{\boldmath$s$}}$, where 
${\mbox{\boldmath$s$}}$ is the neutron 
spin.
The g-factor $\gamma=1.91$.
This magnetic moment at the position ${\bf r}_{N}$
induces
the magnetic field at the position ${\bf r}$ as
$
{\bf B}({\bf{r}}-{\bf{r}}_{N})=\nabla_{{\bf{r}}}\times
\left( {\mbox{\boldmath$\mu$}}_{N}\times
{ {\bf{r}}-{\bf{r} }_{N} \over \mid {\bf{r}}-{\bf{r}}_{N}\mid^{3} } \right).
$
Fluctuating electronic spin at the position ${\bf{r}}_{e}$ interacts with 
this magnetic field. We note that in case of bi-layer compounds,
${\bf{r}}_{e}$ is confined in layer-1 or layer-2. 
This interaction can be expressed by 
$
{\cal H}_{el-neu}=-{(e/ m_{e}c)}
\sum_{{\bf r}_{e}\in {\rm layer-1,2} }{\bf S}({\bf r}_{e},t)\cdot 
{\bf B}({\bf{r}}_{e}-{\bf{r}}_{N}).
$

The initial and final state of the electron-neutron system can be labeled
by
$\mid{\k}_{i}\sigma_{i}E_{i}>$ and $\mid{\k}_{f}\sigma_{f}E_{f}>$
respectively where ${\k}\sigma$ denote a wave number and spin of neutron and $E$ denotes 
a electronic state.
Then   the inelestic neutron scattering cross section can be given by\cite{MarshallLaude}
$$
{d\sigma\over d\Omega d\omega}
=\left({m_{N}\over 2\pi\hbar}\right)^{2}
{k_{f}\over k_{i}}
\sum_{\sigma_{i}\sigma_{f}}\sum_{E_{i}E_{f}}
P_{E_{i}}P_{\sigma_{i}} 
\mid<{\k}_{i}\sigma_{i}E_{i}\mid{\cal H}_{el-neu} 
\mid {\k}_{f}\sigma_{f}E_{f}>\mid^{2}
\delta\left({\rm energy}\right),
$$
where $P_{E_{i}}$ and $P_{\sigma_{i}}$ denote probability factor  
 and the delta function represents energy conservation.
By using the fluctuation-dissipation thoerem, we get 
\begin{equation}
{d\sigma\over d\Omega d\omega}\propto \sum_{\alpha\beta}
(\delta_{\alpha\beta}-\hat{q}_{\alpha}\hat{q}_{\beta})
{{\Im} \chi^{\alpha\beta}_{//}(\q,\omega)+
{\Im} \chi^{\alpha\beta}_{\perp}(\q,\omega)\over 1-\exp(-\omega/T)}.
\label{eqn:neutron cross section}
\end{equation}
From this result, we can see that {\it both of 
intra- and inter-layer spin-spin correlation equally contribute to the inelastic neutron
scattering cross section.}

\begin{flushleft}
\underline{\it NMR Longitudinal  Relaxation Process}
\end{flushleft}

In the case of YBCO, Hamiltonian for the spin ${\bf{I}}_{mi}$ of the $i$-th $^{63}$Cu 
($m$ is a layer index) can be written by
$
{\cal{H}}=-\gamma_{63}{\bf{H}}\cdot({\bf1}+^{63}\!\!{\rm \bf K}^{orb})\cdot{\bf{I}}_{mi}+
{\cal{H}}_{el-nuc},
$
where $\gamma_{63}$ is the g-factor and  $^{63}{\rm \bf K}^{orb}$ 
is the orbital contribution. 
The interaction between 
the $^{63}$Cu nucler spin  and
the neighboring electronic spins can be expressed as
$
{\cal{H}}_{el-nuc}=
\sum_{{\bf{r}}_{j}\in{\rm layer-m}}
I{^{\alpha}_{mi} }^{63}\!\!A^{\alpha\beta}_{ij}S^{\beta}({\bf{r}}_{j}).
$
The hyperfine coupling tensor ${\bf{A}}_{ij}$ is proportional to
the amplitude of the electronic Bloch function at the nuclear site in question.
In case of YBCO,  strong hybridization between the neighboring Cu sites
produces the transfer-hyperfine coupling in the form
$ 
^{63}\!\!A^{\alpha\beta}_{ij}=\delta_{ij}A^{\alpha\beta}+\delta_{i,j+\hat{a}}B
$\cite{MilaRice}.
The first term indicates the core polarization, while the second term indicates 
the transfer hyperfine coupling term that is assumed to be isotropic, where 
$i$ and $j+\hat{a}$ denote the nearest Cu sites.
We note that {\it unless there is the strong hybridyzation between the adjacent layers,
we cannot expect the inter-layer spin correlation contributes to the $T_{1}$}.

Then  the relaxation rate $T_{1}$ can be obtained by
\cite{MoriyaGeneral}
\begin{equation}
{1\over T_{1}}\propto T\sum_{\scriptsize \q}
\mid ^{63}\!\!\!\!A_{\scriptsize \q}\mid^{2} { {\Im}\chi^{-+}_{//}(\q,\omega) \over \omega}
\label{eqn:NMR},
\end{equation}
where $\omega$ is the nuclear Larmor frequency.
The Fourier transform of the hyperfine coupling for YBCO
is given by 
$
^{63}\!\!A_{{\scriptsize \q}//}=A_{//}+2B(\cos q_{x}a+\cos q_{y}a)$,
$^{63}\!\!A_{{\scriptsize \q}\perp}=A_{\perp}+2B(\cos q_{x}a+\cos q_{y}a)$
for $^{63}$Cu sites.
We should stress that {\it only intra-layer spin
fluctuations contribute to the NMR longitudinal relaxation}.
This situation is completely different from the case of the neutron scattering.

\begin{flushleft}
\underline{\it Spin Echo Decay Rate}
\end{flushleft}

The spin echo detects the refocused  transverse spin components 
after appropreate pulses. If  the nuclear spins 
 flip during the experiment, the spin echo decays.
In cuprates, the spin echo decay  arises from mutual spin flips caused by
the indirect interaction
between the nuclear spins via the electronic spin fluctuations.
 This interaction is the same as the well known RKKY 
interaction in  the case of an electron
gas and can be written as
$
{\cal H}_{int}=-(\gamma_{n}\hbar)^{2}\sum_{mn}\left\{
\sum_{{\bf r}_{k}\neq{\bf r}_{l}\in {\rm layer}-m,n}
I{^{\alpha}_{mi}}^{63}A^{\alpha\mu}_{ik}\Re\chi^{\mu\nu} ({\bf r}_{k}-{\bf r}_{l})
^{63}A^{\nu\beta}_{lj}I^{\beta}_{nj}\right\},
$
where 
$\Re\chi^{\alpha\beta} ({\bf r}_{k}-{\bf r}_{l})$ is the static non-local
susceptibility.
Roughly speaking, the Gaussian decay rate mesures the mean time
of nuclear spin flips due to the indirect interaction.

When we  approximate the decay of the NMR spin-echo envelop by a Gaussian
\cite{typeA123 6.98-1,PenningtonSlichter} as
$
e^{-{t^{2}\over2{T_{2G}}^{2}}},
$
the contribution from the intra-layer  correlation
can be written by
\begin{eqnarray}
\left({1\over T_{2G}}\right)^{2}_{//}\propto
\sum_{\scriptsize \q}
{ A_{\scriptsize \q//} ^{2} \Re\chi_{//}^{zz}(\q,0)^{2}}
-\left\{\sum_{\scriptsize \q}
{ A_{\scriptsize \q//} \Re\chi_{//}^{zz}(\q,0)}\right\}^{2},\label{eqn:T2para}
\end{eqnarray}
while the contribution from the inter-layer correlation
can be written by
\begin{eqnarray}
\left({1\over T_{2G}}\right)^{2}_{\perp}\propto
\sum_{\scriptsize \q}
{ A_{\scriptsize \q//} ^{2} \Re\chi_{\perp}^{zz}(\q,0)^{2}}.\label{eqn:T2perp}
\end{eqnarray}
We note that in the inter-layer counterpart it is not
necessary to eliminate 
the contribution from the term with ${\bf {r}}_{k}={\bf r}_{l}$, since the
adjacend layer  are spatially separated.
Thus the observed spin echo decay rate can be written in the form
\begin{equation}
\left({1\over T_{2G}}\right)^{2}=
\left({1\over T_{2G}}\right)^{2}_{//}+\left({1\over T_{2G}}\right)^{2}_{\perp}.
\end{equation}
We note that {\it both of the intra- and inter-layer spin correlation contribute to
the spin echo decay rate.}

\section{Numerical Results and Discussion}
In Fig.11(a) is shown the energy dependence of
the retarded counterpart of the commensurate RPA dynamical 
susceptibility $\chi^{+-}_{ret}(\q^{*},\omega)$.
The retarded counterpart is obtained by analytic countinuation
by using numerical Pade approximation\cite{Pade}.
We fix the chemical potential to $\mu=-0.54$.
We note that we do not treat the chemical potential shift due to the interaction 
in a self consistent manner and as a result the carrier density cannot be fixed.
We set Hubbard interaction $U=3.25t$.
As temperature becomes lower, the peak around $\omega\sim0.46t$  becomes sharp.
We can assign this peak to the dynamical nesting arising from  
the electron-hole excitation energy spectrum
$\varepsilon_{\k+\q^{*}}-\varepsilon_{\k}=4t(\coskx+\cosky)$ at the
momentum transfer $\q^{*}=(\pi/a,\pi/a)$.

In Fig.~11(b), we show the energy dependence
of the imaginary part of the low energy {\it total} susceptibilities, 
$\chi^{+-}_{ret}(\q^{*},\omega)+\Delta\chi^{+-}_{//ret}(\q^{*},\omega)
+\Delta\chi^{+-}_{\perp ret}(\q^{*},\omega)$,
that can be detected by  neutron scattering experiments.
Now  $\chi$ denotes the intra-layer RPA spin fluctuation  defined by 
(\ref{eqn:intraRPA}).
We can see the lower side of the peak is strongly suppressed
as the temperature is lowered. 
We can assign this suppression to dynamical screening due to the developed
inter-layer exchange scattering processes.

Concerning the characterisitc energy scale of the apparant-gap,  
we get   $\omega_{g}\sim 0.4 t$ that is too large to quantitatively 
explain the neutron and NMR experiments. This situation comes from
our simple treatment for the intra-layer spin fluctuations.
In reality  the intra-layer carrier propagator should be renormalized 
by the  strongly developed intra-layer spin flctuations. 
Then as the temperature is decreased,
the loci of the peak
in the dynamnical spin excitation spectrum shifts to lower energy side,
due to the temperature dependent renormalization factor.
So we can say that if we take into account these flctuation effects, 
 the peak moves to the lower energy side and $\omega_{g}$ can also go
down to smaller value.
To see the mechanism of  dynamical screening in detail,
in Fig.~12, we show separately the contributions from the diagram
(a) and (e) in the case of $T=0.008t$.

In our scenario, both of the diagonal and off-diagonal  spin susceptibility is
suppressed due to the dynamical screening effects due to the strongly enhanced
inter-layer exchange scattering processes, and as a result the intra-layer 
RPA dynamical susceptibility that is strongly enhanced 
due to the intra-layer antiferromagnetic spin fluctuations tends to be suppressed.
Thus we can get the apparent gap-like structure in the low energy spin
excitation spectrum.

Next we consider the NMR longitudinal relaxation rates.
As we have shown, in our scenario, the dynamical screening effects leads to the 
strong suppression of  the low energy spin excitations.
Then we can expect that NMR relaxation rates should be suppressed.
We can see from Fig.~11(b) that actually the slope at $\omega\to 0$, namely
$\lim_{\omega\to 0}{\Im \chi(\q^{*},\omega)/ \omega}$, gradually decreases
as the temperature decreases.
However the suppression of the low energy spin excitation spectrum
can  be observed even in the simple RPA results in Fig.~11(a). 
In this case the suppression
is due to the dynamical nesting. 
This situation shows us it is difficult to discuss NMR relaxation only in terms
of the commensurate fluctuations.

To clarify this point we performed the momentum integration of 
the contribution from the  diagram (a) 
with Mila-Rice form factor\cite{MilaRice}.
The results are shown in Fig.~13.
In this case
the contribution arising from the fluctuations with 
$\q\neq\q^{*}=(\pi/a,\pi/a)$ becomes important
since the dynamical nesting effects observed within the intra-layer RPA is
destroyed at $\q\neq\q^{*}$.
As a result  within the intra-layer RPA, $1/T_{1}T$ monotonically increases as the temperature
decreases.
By taking into account the bi-layer effects, however, the gap like structures
survive since the internal momentum  of the T-matrices in the diagram (a) 
is independent
of the external momentum and can be enhanced regardress of the deviation
of the external momentum from the commensurate momentum.
Thus the PSGA behavior can be obtained in the wide region in the momentum
space and $1/T_{1}T$ can detect PSGA in the present scheme.

Finally we comment on the NMR spin echo decay rates in our scheme.
As was mentioned in the previous section, spin echo decay rates
detect the {\it strength} of the indirect interaction.
In this case the negative sign coming from the diagram (e) is meaningless,
and therefore the development of the contribution of the diagram (e) tends
to enhance the spin echo decay rates. As a result, this enhancement cancels
the screening effect arising from the diagram (a).  
Therefore  the spin echo decay rate is expected to be enhanced in the low 
temperature region where the intra-layer spin fluctuations are
strongly developed. 
This situation may be  a clue to understand the reason why $1/T_{1}T$ and $1/T_{2G}$ 
show different temperature dependence in YBCO\footnote{Now we proceed with
numerical calculations to assert this point. We will present the results elsewhere.}.

\section{Concluding Remarks}
In the present paper we proposed a possible mechanism of PSGA in
magnetically coupled bi-layer cuprates.
We shall summarize a physical picture for the pseudo-spin gap formation in the present scheme. 
The basic steps are as follows.\\

\noindent
STEP 1: In a lightly doped bi-layer cuprate, an intra-layer itinerant electronic  system 
lies just near the antiferromagnetc phase boundary. \\

\noindent
STEP 2: The itineracy of the intra-layer system dramatically enhances inter-layer antiferromagnetc
coupling. As a result inter-layer coupling becomes temperature dependent and
developes dramatically as the temperature decreaces.\\

\noindent
STEP 3: Just near the magnetic phase boundary
the inter-layer particle-hole ladder becomes very important.
Dramatically enhanced inter-layer antiferromagnetic coupling induces inter-layer
exchange scattering processes. 
This process leads to strong attraction for
inter-layer spin singlet pair and strong repulsion for inter-layer 
spin triplet pair. Namely the strong repulsion between  parallel spins is
strongly enhanced
as the intra-layer system goes nearer to the magnetic phase boundary.\\

\noindent
STEP 4: Strong repulsion between parallel spins 
lead to  the dynamical screening
of the total magnetic excitations in the low energy region.
Thus PSGA can be realized in a magnetically coupled bi-layer cuprate.

In the present work, we have not payed our attention to
the self consistent renormalization effects.
Now we note that the inter-layer exchange scattering processes do not
contribute to any vertex correction, since these processes do not
affect the number conservation of carriers. 
Thus it is sufficient to consider the self-energy effects.
When we consider the self consistency,
the diagram (a) and (b)  in Fig.~9
should be regarded as the self energy diagram.
Then the bare green function ${\cal{G}}_{0}(k)$ should be replaced by
the dressed one
$
{\cal{G}}(k)=[i\varepsilon_{n}-\xi_{\k}-\Sigma(k)]^{-1},
$
where the self energy is given by
$
\Sigma(k)={T\over N}\sum_{Q}
[{\cal{T}}^{\up\up,\up\up}_{\perp FW}(Q)+{\cal{T}}^{\up\down,\up\down}_{\perp FW}(Q)]{\cal{G}}(k+Q).
$
By putting 
$
x(k)=\Re \Sigma(k)$,
$z(k)=1-{\Im \Sigma(k)\over \varepsilon_{n}}$,
we can write the green function in a form
$
{\cal G}(k)={1\over z(k)}{1\over i\varepsilon_{n}-\tilde\xi_{\footnotesize\k}},
$
where
$
\tilde\xi_{\footnotesize\k}={\xi_{\footnotesize\k}+x(k)\over z(k)}.
$
Then the irreducible loop in RPA is replaced by
$
\tilde\chi(q)
=-{1\over 2}\sum_{\footnotesize\k}{1\over z(k) z(k+q)}
{\tanh{\tilde\xi_{\footnotesize\k+\q}/ 2T}-\tanh{\tilde\xi_{\footnotesize\k}/ 2T}
\over i\omega_{l}-\tilde\xi_{\footnotesize\k+\q}+\tilde\xi_{\footnotesize\k}}.
$
In this case we can expect that the dynamical screening effects in the present context
corresponds to
the enhancement of mass-renormalization factor $z(k)$ near the antiferromagnetic phase boundary.
We may say that in our scenario the mass-renormalization depends sensitevely on
$\omega$ and the low energy mass becomes heavier than the high energy mass.
If this mechanism really takes place, PSGA is  obtained again.
We will report the results of  sophisticated calculations in the forthcoming paper.


\section*{Acknowledgements}

I would like to express my  thanks to Professors H.Namaizawa,
M.Ogata and H.Fukuyama for valuable discussions
and critical comments.
I gratefully acknowledge helpful discussion with Dr.Y.Itoh on 
experimental insights, especially on NMR experiments.
Furthermore I would like to thank to Dr.Shiina for useful advice
on numerical work.

\section*{Appendix: Spin-rotational  
Symmetry  Relation}
Now we show the spin rotational
invariance relation in the present scheme
 by taking into account all the diagrams for the contribution
to the longitudinal susceptibility.
Necessary diagrams are shown in  Fig.~14. 
Our goal is to show the relation 
\begin{eqnarray}
\Delta\chi^{+-}_{//}(q)&=&2\Delta\chi^{zz}_{//}(q),\\
\Delta\chi^{+-}_{\perp}(q)&=&2\Delta\chi^{zz}_{\perp}(q),\label{eqn:SRIRMM}
\end{eqnarray}
where 
$$
\Delta\chi^{zz}_{mn}(q)={1\over 4}\sum_{\sigma}
[\Delta\chi^{\sigma,\sigma}_{mn}(q)-
\Delta\chi^{\sigma,-\sigma}_{mn}(q)].
$$
Contribution to the longitudinal diagonal susceptibility  can be obtained 
through the
diagrammatic algebra shown in Fig.~15(a) and we get
\begin{eqnarray}
\Delta\chi^{\sigma,\sigma}_{//}(q)
-\Delta\chi^{\sigma,-\sigma}_{//}(q)=
\gamma(q)^{2}{T\over N}\sum_{Q}
\{{\cal{T}}^{\up\up,\up\up}_{\perp FW}(Q)+
{\cal{T}}^{\up\down,\up\down}_{\perp FW}(Q)\}{\cal L}^{(1)}(q,Q)\nonumber\\
+\gamma(q)^{2}{T\over N}\sum_{Q}\{
{\cal{T}}^{\up\up,\up\up}_{\perp FW,BW}(Q){\cal{T}}^{\up\up,\up\up}_{\perp FW,BW}(Q+q)+\nonumber\\
{\cal{T}}^{\up\down,\up\down}_{\perp FW,BW}(Q){\cal{T}}^{\up\down,\up\down}_{\perp FW,BW}(Q+q)
-{\cal{T}}^{\up\down,\down\up}_{\perp FW,BW}(Q)
{\cal{T}}^{\up\down,\down\up}_{\perp FW,BW}(Q+q)\}
\{{\cal L}^{(3)}(q,Q)\}^{2},\label{eqn:mmlond}
\end{eqnarray}
where 
\begin{eqnarray*}{\cal{T}}^{\mu\nu,\rho\lambda}_{\perp FW,BW}(Q)
{\cal{T}}^{\mu'\nu',\rho'\lambda'}_{\perp FW,BW}(Q+q)\equiv 
{\cal{T}}^{\mu\nu,\rho\lambda}_{\perp FW}(Q)
{\cal{T}}^{\mu'\nu',\rho'\lambda'}_{\perp FW}(Q+q)
+{\cal{T}}^{\mu\nu,\rho\lambda}_{\perp BW}(Q)
{\cal{T}}^{\mu'\nu',\rho'\lambda'}_{\perp BW}(Q+q).\end{eqnarray*}
Furthermore  the algebra shown in Fig.~15(b) produces
 the off-diagonal counterpart as
\begin{eqnarray}
\Delta\chi^{\sigma\sigma}_{\perp}(q)
-\Delta\chi^{\sigma,-\sigma}_{\perp}(q)=
-         \gamma(q)^{2}
{T\over N}\sum_{Q}
\{{\cal{T}}^{\up\up,\up\up}_{\perp BW}(Q)-{\cal{T}}^{\up\down,\up\down}_{\perp BW}(Q)\}
 {\cal L}^{(2)}(q,Q)
\nonumber\\
+         \gamma(q)^{2}
{T\over N}\sum_{Q}
\{{\cal{T}}^{\up\up,\up\up}_{\perp FW,BW}(Q){\cal{T}}^{\up\up,\up\up}_{\perp FW,BW}(Q+q)
-\nonumber\\
{\cal{T}}^{\up\down,\up\down}_{\perp FW,BW}(Q){\cal{T}}^{\up\down,\up\down}_{\perp BW,FW}(Q+q)-
{\cal{T}}^{\up\down,\down\up}_{\perp FW,BW}(Q){\cal{T}}^{\up\down,\down\up}_{\perp BW,FW}(Q+q)
\}{\cal L}^{(3)}(q,Q){\cal L}^{(3)}(-q,-Q),\label{eqn:mmlono}
\end{eqnarray}
where 
\begin{eqnarray*}{\cal{T}}^{\mu\nu,\rho\lambda}_{\perp FW,BW}(Q)
{\cal{T}}^{\mu'\nu',\rho'\lambda'}_{\perp BW,FW}(Q+q)\equiv 
{\cal{T}}^{\mu\nu,\rho\lambda}_{\perp FW}(Q)
{\cal{T}}^{\mu'\nu',\rho'\lambda'}_{\perp BW}(Q+q)
+{\cal{T}}^{\mu\nu,\rho\lambda}_{\perp BW}(Q)
{\cal{T}}^{\mu'\nu',\rho'\lambda'}_{\perp FW}(Q+q).\end{eqnarray*}
From now on we omitt the indices {\it FW} and {\it BW}, since these
indices are not essential for the spin rotational symmetry.
By noting the spin rotaional invariance relation for the exchange 
scattering T-matrixes, (\ref{eqn:SRS2}),
$$
{\cal{T}}^{\up\up,\up\up}_{\perp}(Q)-{\cal{T}}^{\up\down,\down\up}
_{\perp}(Q)={\cal{T}}^{\up\down,\up\down}_{\perp}(Q),
$$
 we can rewrite the terms in the braces in (\ref{eqn:mmlond}) as,
\begin{eqnarray*}
{\cal{T}}^{\up\up,\up\up}_{\perp}(Q){\cal{T}}^{\up\up,\up\up}_{\perp}(Q+q)+
{\cal{T}}^{\up\dwon,\up\down}_{\perp}(Q){\cal{T}}^{\up\dwon,\up\down}_{\perp}(Q+q)
-{\cal{T}}^{\up\down,\down\up}_{\perp}(Q){\cal{T}}^{\up\down,\down\up}_{\perp}(Q+q)\\
={\cal{T}}^{\up\up,\up\up}_{\perp}(Q){\cal{T}}^{\up\up,\up\up}_{\perp}(Q+q)+
{\cal{T}}^{\up\dwon,\up\down}_{\perp}(Q){\cal{T}}^{\up\dwon,\up\down}_{\perp}(Q+q)\\
-\{{\cal{T}}^{\up\up,\up\up}_{\perp}(Q)-{\cal{T}}^{\up\dwon,\up\down}_{\perp}(Q)\}
\{{\cal{T}}^{\up\up,\up\up}_{\perp}(Q+q)-{\cal{T}}^{\up\dwon,\up\down}_{\perp}(Q+q)\}\\
=2\{{\cal{T}}^{\up\up,\up\up}_{\perp}(Q){\cal{T}}^{\up\dwon,\up\down}_{\perp}(Q+q)
+{\cal{T}}^{\up\dwon,\up\down}_{\perp}(Q){\cal{T}}^{\up\up,\up\up}_{\perp}(Q+q)\},
\end{eqnarray*}
and similarly in (\ref{eqn:mmlono}),
\begin{eqnarray*}
{\cal{T}}^{\up\up,\up\up}_{\perp}(Q){\cal{T}}^{\up\up,\up\up}_{\perp}(Q+q)-
{\cal{T}}^{\up\dwon,\up\down}_{\perp}(Q){\cal{T}}^{\up\dwon,\up\down}_{\perp}(Q+q)
-{\cal{T}}^{\up\down,\down\up}_{\perp}(Q){\cal{T}}^{\up\down,\down\up}_{\perp}(Q+q)\\
=\{{\cal{T}}^{\up\down,\down\up}_{\perp}(Q+q)+{\cal{T}}^{\up\dwon,\up\down}_{\perp}(Q+q)\}
\{{\cal{T}}^{\up\down,\down\up}_{\perp}(Q)+{\cal{T}}^{\up\dwon,\up\down}_{\perp}(Q)\}\\-
{\cal{T}}^{\up\dwon,\up\down}_{\perp}(Q){\cal{T}}^{\up\dwon,\up\down}_{\perp}(Q+q)
-{\cal{T}}^{\up\down,\down\up}_{\perp}(Q){\cal{T}}^{\up\down,\down\up}_{\perp}(Q+q)\\
=2\{{\cal{T}}^{\up\down,\down\up}_{\perp}(Q){\cal{T}}^{\up\dwon,\up\down}_{\perp}(Q+q)
+{\cal{T}}^{\up\dwon,\up\down}_{\perp}(Q)
{\cal{T}}^{\up\down,\down\up}_{\perp}(Q+q)\}
\end{eqnarray*}
Thus we get the results 
\begin{eqnarray*}
\Delta\chi^{\sigma,\sigma}_{//}(q)
-\Delta\chi^{\sigma,-\sigma}_{//}(q)=\nonumber\\
\gamma(q)^{2}{T\over N}\sum_{Q}
\{{\cal{T}}^{\up\up,\up\up}_{\perp FW}(Q)+
{\cal{T}}^{\up\down,\up\down}_{\perp FW}(Q)\}{\cal L}^{(1)}(q,Q)\nonumber\\
2\gamma(q)^{2}{T\over N}\sum_{Q}
\{{\cal{T}}^{\up\up,\up\up}_{\perp}(Q){\cal{T}}^{\up\dwon,\up\down}_{\perp}(Q+q)
+{\cal{T}}^{\up\dwon,\up\down}_{\perp}(Q){\cal{T}}^{\up\up,\up\up}_{\perp}(Q+q)\}
\{{\cal L}^{(3)}(q,Q)\}^{2},
\end{eqnarray*}
and 
\begin{eqnarray*}
\Delta\chi^{\sigma\sigma}_{\perp}(q)
-\Delta\chi^{\sigma,-\sigma}_{\perp}(q)=
-2         \gamma(q)^{2}
{T\over N}\sum_{Q}
{\cal{T}}^{\up\down,\down\up}_{\perp}(Q)
 {\cal L}^{(1)}(q,Q)\nonumber\\
+2         \gamma(q)^{2}
{T\over N}\sum_{Q}
\{{\cal{T}}^{\up\down,\down\up}_{\perp}(Q)
{\cal{T}}^{\up\dwon,\up\down}_{\perp}(Q+q)\nonumber\\
+{\cal{T}}^{\up\dwon,\up\down}_{\perp}(Q){\cal{T}}^{\up\down,\down\up}_{\perp}(Q+q)\}
{\cal L}^{(3)}(q,Q){\cal L}^{(3)}(-q,-Q),
\end{eqnarray*}
These expressions are exactly twice the corresponding transverse susceptibility, 
(\ref{eqn:correctionparallel})
and (\ref{eqn:correctionparallel}). Thus we obtain the spin rotational symmetry relation (\ref{eqn:SRIRMM}). 

\pagebreak

\pagebreak
\section*{Figure Captions}
\baselineskip14pt
\noindent
Fig.~1: Magnetic unit cell of YBa$_{2}$Cu$_{3}$O$_{6}$.
Filled and open  circles denote antiparalle spins at  planar Cu$^{2+}$ sites, 
while shaded circles denote nonmagnetic 
Cu$^{1+}$ ions. 

\bigskip
\noindent
Fig.~2: Fundamental processes induced by the inter-layer
magnetic interaction;\\
(a) The scattering process  with spin flip (type-A), \\ 
(b) the scattering processes between parallel spins (type-B) and \\
(c) the scattering processes between anti-parallel spins (type-C).

The straight  and  wavy line represent respectively 
the green's function of an  itinerant carrier and 
the inter-layer antiferromagnetic interaction.
Here the thick and thin line correspond to an carrier in the layer-1
and the layer-2 respectively.  

\bigskip
\noindent
Fig.~3: Enhancement of the inter-layer exchange interaction vertices
due to the intra-layer RPA spin fluctuations.\\
(a): The intra-layer
transverse spin fluctuations enhances the Type-A scattering vertex.  \\
(b), (c): The intra-layer
longitudinal spin fluctuations enhances both of the Type-B and Type-C scattering 
vertex.
The straight  and  wavy line represent respectively 
the green's function of an itinerant carrier and 
the inter-layer antiferromagnetic interaction.
The dotted line represents the intra-layer Hubbard interaction.
   
\bigskip
\noindent
Fig.~4: These figures show how the intra-layer spin fluctuations
in different layer couple to each other via the inter-layer Heisenberg
interaction.
The intra-layer transverse spin fluctuations can couple to 
each other via the type-a inter-layer scattering channel,  as is shown 
in (a).
On the other hands the intra-layer longitudinal spin fluctuations
can couple to each other via both of the type-b and type-c inter-layer
scattering channel,  as is shown in (b) and (c). 
The shaded triangle in the right hand side represents the vertex enhancement due to
the intra-layer spin fluctuations where  $\gamma(q)$ is the Stoner factor.
We can see that the inter-layer interaction is enhanced through the {\it double}
Stoner factor and strongly enhanced.

\bigskip
\noindent
Fig.~5: The  $n$-th order scattering processes for three scattering 
channels corresponding to the
different spin dependent scattering channel.\\
(a) the  T-matrix that represents the electron-hole 
exchange scattering process between untiparallel spin, \\
(b) the   T-matrix that represents the electron-hole 
exchange scattering process between parallel spin, 
and\\
(c) the auxiliary T-matrix that consists of only B-type scattering channel.

\bigskip
\noindent
Fig.~6: Restrictions coming from the momentum tansfer.
Since the triangle vertex is enhanced around the momentum transfer 
{\bf{q}} $\sim$ {\bf{q}}$^{*}$,  
the incoming momentum({\bf{q}}) and outgoing momentum({\bf{k}}$'$) 
of an itinerant carrier 
are restricted.\\
Case (a): When 
{\bf{k}}$-${\bf{k}}$'\sim$ {\bf{q}}$^{*}$ (backward scattering), only the processes
 including odd number of  triangles are enhanced.\\
Case (b): When {\bf{k}}$-${\bf{k}}$''\sim 0$ (forward scattering),  only the processes including even number of  
triangles are enhanced.

\bigskip
\noindent
Fig.~7:
Behavior of T-matrices in the singlet and triplet channel
as a function of 
${3\over 4}{\cal{J}}_{\perp}(T)\chi_{0}$. 
Each line represents the forward processes  and 
backward processes distinctively.

\bigskip
\noindent
Fig.~8: The temperature dependence of the averaged inter-layer 
interaction ${\cal{J}}_{\perp}(T)$ for $U=3.25 t$  and  $J_{0\perp}=0.08t$
for the chemical potential $\mu=-0.54$.

\bigskip
\noindent
Fig.9: The  contributions to the transverse
susceptibility arising from the  enhanced inter-layer exchange scattering process.
The graph (a),  (b), (c), (d) contribute to the diagonal susceptibility 
and the graph (e), (f), (g) contribute to the off-diagonal susceptibility.
The straight and broken lines represent the propagator of the carrier in  different 
layers.

\bigskip
\noindent
Fig.~10: The  behavior of the  integrand in the coupling function
${\cal L}^{(1)}_{\footnotesize{\mbox{\boldmath$k$}}}(\q^{*};i\omega_{l};{\mbox{\boldmath$q$}}^{*};i\Omega_{m}=0)$
in the first quadrant of the Brillouin zone
at  temperature $T=0.06t$ and $T=0.01t$.
We set $l=1$ and $\mu=-0.54t$.

\bigskip
\noindent
Fig.~11:
(a) Imaginary parts of the intra-layer RPA susceptibility
for various temperatures $T$.\\
(b)The retarded counterpart of the total susceptibilities 
$\chi^{+-}_{\rm ret}({\mbox{\boldmath$q$}}^{*};\omega)$ for various temperatures $T$.
We fixed the chemical potential to $\mu=-0.54$.
We set  $U=3.25 t$  and  $J_{0\perp}=0.08t$.

\bigskip
\noindent
Fig.~12: The retarded counterpart of the  dynamical susceptibility.
We show contributions from diagram (a) and (e) separately.
[RPA+Diagram(a)] gives the diagonal susceptibility that contribute to
NMR longitudinalrelaxation.
[RPA+Diagram(a), (e)] is the sum of diagonal and off-diagonal susceptibility
that contribute to the neutron scattering experiments.

\bigskip
\noindent
Fig.~13:
NMR longitudinal relaxation rate.
The dashed line represents the result from intra-layer RPA, 
while the solid line represents the result when the
inter-layer effects are taken into account.

\bigskip
\noindent
Fig.~14: Diagrams that are necessary to guarantee the spin rotaional
invariance relation when we include the  coupling
arising from enhanced inter-layer exchange scattering processes;\\
(a) for the diagonal process,  and\\
(b) for the off-diagonal process.
\end{document}